\documentclass[sigconf]{acmart}
\usepackage{mathtools}
\usepackage{amsthm}
\usepackage{subcaption}
\usepackage{paralist}
\usepackage{bm}
\usepackage{amsfonts}
\usepackage{multirow}
\usepackage{cases}
\usepackage{booktabs}
\usepackage{threeparttable}
\usepackage{epstopdf}
\usepackage{upgreek}
\usepackage{endnotes}
\usepackage{etoolbox}
\usepackage{algpseudocode}
\usepackage{amsmath}
\usepackage{algorithm}
\usepackage{pythonhighlight}
\usepackage{xspace}
\usepackage{array}
\usepackage{enumitem}

\AtBeginDocument{%
  \providecommand\BibTeX{{%
    \normalfont B\kern-0.5em{\scshape i\kern-0.25em b}\kern-0.8em\TeX}}}
\newcommand{\eat}[1]{}

\newcommand{\ie}{\emph{i.e.,}\xspace}
\newcommand{\eg}{\emph{e.g.,}\xspace}

\copyrightyear{2020}
\acmYear{2020}
\setcopyright{acmcopyright}\acmConference[SIGIR '20]{Proceedings of the 43rd International ACM SIGIR Conference on Research and Development in Information Retrieval}{July 25--30, 2020}{Virtual Event, China}
\acmBooktitle{Proceedings of the 43rd International ACM SIGIR Conference on Research and Development in Information Retrieval (SIGIR '20), July 25--30, 2020, Virtual Event, China}
\acmPrice{15.00}
\acmDOI{10.1145/3397271.3401043}
\acmISBN{978-1-4503-8016-4/20/07}

\begin{document}
\fancyhead{}
\title{ESAM: Discriminative Domain Adaptation with Non-Displayed Items to Improve Long-Tail Performance}\thanks{$^{\ddagger}$ This work was done as a research intern in Alibaba Group.}\thanks{$\star$ Rong Xiao contributed equally to this research.}\thanks{$\dagger$ Chenliang Li is the corresponding author.}



\author{Zhihong Chen$^{12\ddagger}$, Rong Xiao$^{2\star}$, Chenliang Li$^{3\dagger}$, Gangfeng Ye$^2$, Haochuan Sun$^2$, Hongbo Deng$^2$}
\affiliation{
\institution{
	$^1$Institute of Information Science and Electronic Engineering, Zhejiang University, Hangzhou, China\\
$^2$Alibaba Group, Hangzhou, China\\
$^3$School of Cyber Science and Engineering, Wuhan University, Wuhan, China\\
$^1$zhihongchen@zju.edu.cn, $^2$\{xiaorong.xr, gangfeng.ygf, haochuan.shc, dhb167148\}@alibaba-inc.com\\
$^3$cllee@whu.edu.cn
}
}
\def\authors{Zhihong Chen, Rong Xiao, Chenliang Li, Gangfeng Ye, Haochuan Sun, Hongbo Deng}

%
%
%
%

\renewcommand{\shortauthors}{Trovato and Tobin, et al.}

\begin{abstract}
Most of ranking models are trained only with displayed items (most are hot items), but they are utilized to retrieve items in the entire space which consists of both displayed and non-displayed items (most are long-tail items). Due to the sample selection bias, the long-tail items lack sufficient records to learn good feature representations, \ie data sparsity and cold start problems. The resultant distribution discrepancy between displayed and non-displayed items would cause poor long-tail performance. To this end, we propose an entire space adaptation model (ESAM) to address this problem from the perspective of domain adaptation (DA). ESAM regards displayed and non-displayed items as source and target domains respectively. Specifically, we design the attribute correlation alignment that considers the correlation between high-level attributes of the item to achieve distribution alignment. Furthermore, we introduce two effective regularization strategies, \ie \textit{center-wise clustering} and \textit{self-training} to improve DA process. Without requiring any auxiliary information and auxiliary domains, ESAM transfers the knowledge from displayed items to non-displayed items for alleviating the distribution inconsistency. Experiments on two public datasets and a large-scale industrial dataset collected from Taobao demonstrate that ESAM achieves state-of-the-art performance, especially in the long-tail space. Besides, we deploy ESAM to the Taobao search engine, leading to significant improvement on online performance. The code is available at \url{https://github.com/A-bone1/ESAM.git}
\end{abstract}



\maketitle
\section{Introduction}
A typical formulation of the ranking model is to provide a rank list of items given a query. It has a wide range of applications, including recommender systems \cite{koren2009matrix,rendle2009bpr}, search systems \cite{burges2010ranknet,joachims2002optimizing}, and so on. The ranking model can be formulated as: $q \xrightarrow{R} \hat{\mathbb{D}}_{q}$, where $q$ is the query, e.g., user profiles and user behavior history in the recommender systems, and user profiles and keyword in the personalized search systems, depending on specific rank applications. $\hat{\mathbb{D}}_{q}$ represents the rank list of related items (e.g., textual documents, information items, answers) retrieved based on $q$,  and $R=\{r_{i}\}_{i=1}^{n_{d}}$ consists of the relevance scores $r_{i}$ between $q$ and each item $d_{i}$ in the entire item space, $n_{d}$ is the total number of items. In short, the ranking model aims to select the top K items with the highest relevance to the query as the final ranking result.

\begin{figure}
  \centering
  \includegraphics[width=0.45\textwidth]{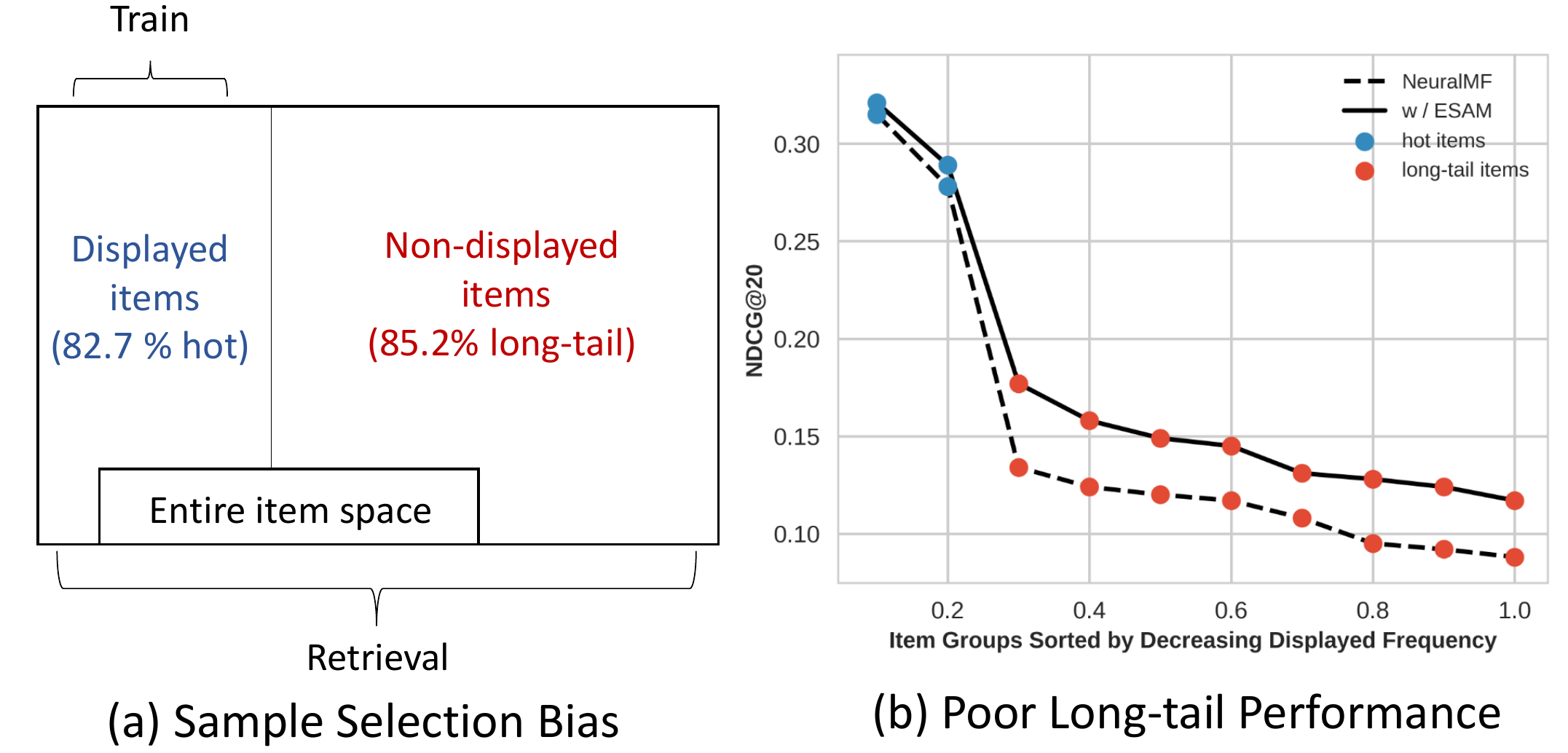}
  \caption{(a) The SSB problem. (b) The SSB cause poor long-tail performance on CIKM Cup 2016 dataset.}
  \label{figv21}
\end{figure}

\begin{figure*}
  \includegraphics[width=0.8\textwidth]{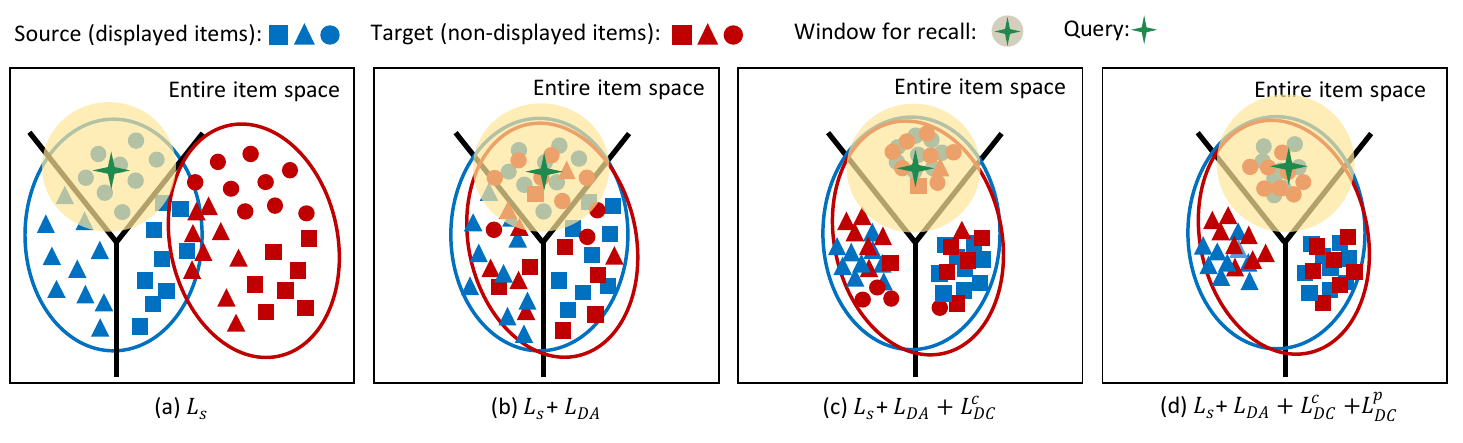}
  \caption{Three shapes represent three feedbacks, the star with a shadow represents a query feature with its retrieval range. (a) Domain shift between displayed and non-displayed item features. (b) Although the domain shift is alleviated, the poor neighborhood relationship makes the model easy to retrieve outliers. (c) The proposed center-wise clustering encourages better manifold structure. However, the class-agnostic ACA may cause negative transfer, \eg some target circles are aligned to source triangles. (d) ESAM extracts the best item feature representations.}
  \label{fig:teaser}
\end{figure*}

Currently, deep learning-based methods are widely used in the ranking models. For example, \cite{cheng2016wide,li2019multi,zhu2019improving,sigir20:he} in the application of recommender systems, \cite{burges2005learning,huang2013learning} in the application of search systems, and \cite{severyn2015learning} in the application of text retrieval. These methods show a better ranking performance than traditional algorithms \cite{sarwar2001item,singh2008relational}. However, these models are mainly trained using only the implicit feedbacks (e.g., clicks and purchases) of the displayed items, but utilized to retrieval items in the entire item space including both displayed and non-displayed items when providing services. We divide the entire item space into hot items and long-tail items by the display frequency. By analyzing two public datasets (MovieLens and CIKM Cup 2016 datasets), we found that 82\% of displayed items are hot items, while 85\% of non-displayed items are long-tail items. Therefore, as shown in Fig.~\ref{figv21}a, the existence of sample selection bias (SSB)\cite{yuan2019improving} causes the model to overfit the displayed items (most are hot items), and cannot accurately predict long-tail items (Fig.~\ref{figv21}b). What's worse,  this training strategy makes the skew of the model towards popular items \cite{krishnan2018adversarial}, which means that these models usually retrieve hot items while ignoring those long-tail items that would be more suitable, especially those that are new arrival. This phenomenon is called  "Matthew Effect" \cite{liu2019real}.  We believe the reason for this phenomenon is that the SSB leads to the lack of records (\ie feedbacks) of long-tail items to obtain good feature representations, so the features of long-tail items have an inconsistent distribution compared to that of hot items whose records are adequate.  As shown in Fig.~\ref{fig:teaser}a, the existence of domain shift \cite{candela2009dataset} means that these ranking models are difficult to retrieve long-tail items, because they always overfit the hot items.

To improve the long-tail performance of ranking models, and increase the diversity of retrieval results, existing methods exploit auxiliary information \cite{yuan2019darec,yuan2019improving} or auxiliary domains \cite{gao2019cross,man2017cross}, which are not easily accessible.  For example, \cite{gao2019cross,kanagawa2019cross} use two sample spaces (\eg MovieLens and Netflix) to implement knowledge transfer through shared items, and \cite{yuan2019improving} uses a randomly displayed unbiased dataset to fine-tune the model. Given the diversity of model architectures and applications \cite{he2017neural,liang2018variational}, architectural solutions may not generalize well. By following some past works \cite{lakkaraju2017selective}, we highlight the importance of \textit{learning good feature representations} for non-displayed items. To achieve this, through considering poor long-tail performance is caused by the domain shift between displayed and non-displayed items and the fact that non-displayed items are unlabeled instances,  we adopt unsupervised domain adaptation (DA) technology, which regards displayed and non-displayed items as source and target domains respectively. The DA method allows the application of the model trained with labeled source domain to target domain with limited or missing labels. \eat{Hence, we can alleviate the distribution shift caused by a lack of sufficient records of long-tail items.}

Previous DA-based works reduce the domain shift by minimizing some measures of distribution, such as maximum mean discrepancy (MMD) \cite{tran2019domain}, or with adversarial training \cite{krishnan2018adversarial}. For the ranking task, we propose a novel DA method, named \textit{attribute correlation alignment} (ACA). Whether an item is displayed or not, the correlation between its attributes follows the same rules (knowledge). For example, in the e-commerce, the more luxurious the brand of an item, the higher the price (brand and price are item attributes). This rule is the same for displayed and non-displayed items. In a ranking model, each item will be represented as a feature representation through feature extractor, and each dimension of the feature can be regarded as a high-level attribute of the item. Therefore, we believe that the correlation between high-level attributes should follow the same rules in both the displayed and non-displayed spaces. However, the model cannot get the features of non-displayed items well due to the lack of labels,  which causes the inconsistency between displayed and non-displayed item feature distributions and makes the above paradigm untenable.  Hence, we design the attribute correlation congruence (A2C) to exploit the pair-wise correlation between high-level attributes as distribution.

Although the aforementioned ACA can deal with the distribution inconsistency, there are two critical limitations: (1) learning with the point-wise cross-entropy ignores the spatial structure information, which leads to poor neighborhood relationship in the feature space \cite{zhu2019improving} (Fig.~\ref{fig:teaser}b); (2) the target label for non-displayed item is not available, which could easily causes the negative transfer \cite{pan2009survey} (Fig.~\ref{fig:teaser}c) when aligning distributions blindly. Negative transfer is a dilemma that the transfer model performs even worse than non-adaptation model \cite{pan2009survey}. To address these two obstacles, we propose two novel regularization strategies, \textit{center-wise clustering} and \textit{self-training}, to enhance DA process. We observed that for a query, items with the same feedback are similar, and items with different feedbacks are dissimilar. For example, in the e-commence, when displaying a variety of mobile phones to a user, the user may click all iPhones and ignore others.  Therefore, for each query, we can categorize displayed items by the feedback type. The proposed center-wise clustering is devised to make similar items cohere together while dissimilar items separate from each other. This constraint could provide further guidance for domain adaptation, leading to better ranking performance. As to the missing of target label, we assign pseudo-labels of high confidence to target items, and make the model fit these items by self-training. Moreover, when taking into account these pseudo-labels while performing alignment, the model can gradually predict more complex items correctly.
In summary, the contributions of this work can be listed as follows: (1) We propose a general entire space adaptation model (ESAM) for ranking models, which exploits the domain adaptation with attribute correlation alignment to improve long-tail performance. ESAM can be easily integrated into most existing ranking frameworks; (2) Two novel yet effective regularization strategies are introduced to optimize the neighborhood relationship and handle the missing of target label for discriminative domain adaptation; (3) We realise ESAM in two typical rank applications: item recommendation and personalized search system. The results on two public datasets, and an industrial dataset collected from Taobao prove the validity of ESAM.  Besides, we deploy ESAM to the Taobao search engine, which also produces better performance through an online A/B testing.

\section{Proposed Method} \label{method}
\begin{figure}
  \centering
  \includegraphics[width=0.40\textwidth]{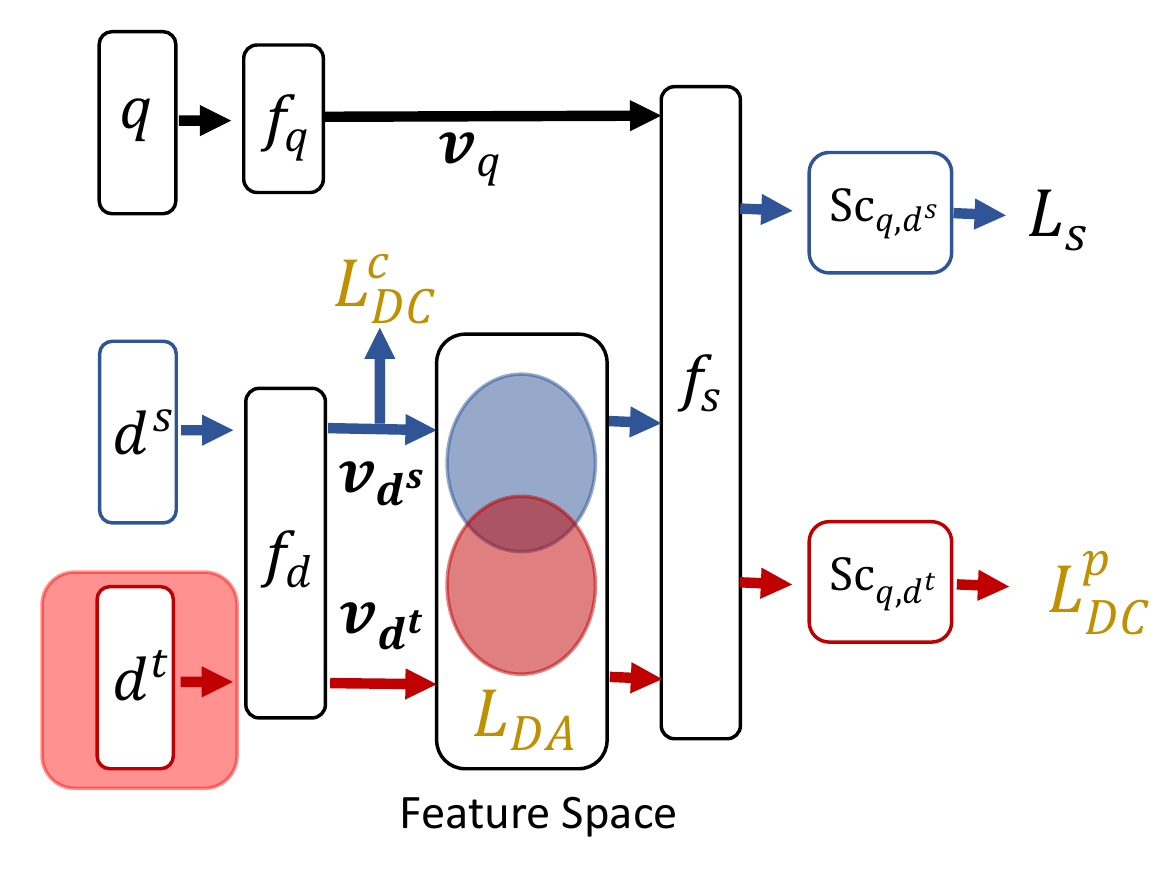}
  \caption{Overview of the ESAM. The red flow represents the non-displayed item stream.}
  \label{figv31}
\end{figure}

In this section, we first briefly introduce the basic ranking framework named BaseModel. Then, ESAM, which contains the proposed A2C and two regularization strategies, is integrated into the BaseModel for better item feature representation learning in the entire space. Fig.~\ref{figv31} shows the overall framework of ESAM. We consider ESAM without the unlabeled non-displayed item input stream (the red flow in the Fig.~\ref{figv31}) as the BaseModel.

\subsection{Preliminaries} \label{Preliminaries}
In this paper, the source domain (displayed items) is denoted as $\mathbb{D}^{s}$, the target domain (non-displayed items) is denoted as $\mathbb{D}^{t}$, and the entire item space is denoted as $\mathbb{D}= \mathbb{D}^{s} \cup \mathbb{D}^{t}$.
Here, the source domain and target domain share the same query set $\mathbb{Q}$. The set of feedback from a query to an item is denoted as $\mathbb{O}=\{q,d,y\}$,  where $q$ is a specific query (e.g., keywords, user profiles, questions, depending on the specific ranking application), d is an item (e.g., textual documents, information items, answers), and $y$ is an implicit feedback. For each $q$, we assign a labeled source item set $\mathcal{D}^s_q=\{(d_{j}^{s},y_{j}^{s})\}_{j=1}^{n}$ whose feedback is available, and an unlabeled target item set $\mathcal{D}^t_q=\{d_{j}^{t}\}_{j=1}^{n}$ randomly selected from non-displayed items. The goal of the ranking model is retrieving a ranked item set $\hat{\mathbb{D}}_{q}$ from $\mathbb{D}$ to maximize query's satisfaction.

\eat{
Note that, we assign each query 10 source items $\mathcal{D}^s_q$ and 10 target items $\mathcal{D}^t_q$.
}
\subsection{BaseModel}
As shown in Fig.~\ref{figv31}, the ranking model mainly consists of a query side $f_q$ and an item side $f_d$. The core task is to map a query $q$ and an item $d$ into a query feature $\bm{v}_q\in\mathbb{R}^{1\times L}$ and an item feature $\bm{v}_d\in\mathbb{R}^{1\times L}$ via $f_q$ and $f_d$ respectively, which can be formulated as:
\begin{equation} \label{fu}
\begin{split}
\bm{v}_q&=f_{q}(q),
\end{split}
\end{equation}
\begin{equation} \label{fi}
\begin{split}
\bm{v}_d&=f_{d}(d).
\end{split}
\end{equation}

When query and item features are extracted, a scoring function $f_s$ is utilized to calculate the relevance score for a query and an item, which can be formulated as:
\begin{equation}\label{score}
\begin{split}
Sc_{q,d}=f_{s}(\bm{v}_q,\bm{v}_d).
\end{split}
\end{equation}

Then, in the training stage, the ranking model can be trained with $L_{s}$ (\eg point-wise objective function \cite{li2008mcrank,nallapati2004discriminative}, pair-wise objective function \cite{burges2005learning}, list-wise objective function \cite{burges2010ranknet,cao2007learning}) by using the feedback. In the serving stage, the rank list of the top K items with the highest relevance score can be used as the retrieval result. Note that the implicit feedbacks are only available in most real-world applications. Thus the point-wise objection loss is widely adopted for model training.

Recently,  many works design novel $f_q$ and $f_d$ to extract better features. For example,  neural matrix factorization (NeuralMF) \cite{huang2013learning} adopts MLP as $f_q$ and $f_d$ to extract query and item features respectively, deep interest evolution network (DIN)~\cite{zhou2019deep} introduces the attention mechanism between $f_q$ and $f_d$, and behavior sequence transformer (BST)~\cite{chen2019behavior} employs the transformer module in $f_q$ to capture short-term interest of query. However, most methods ignore the handling of non-displayed items and these architectural solutions may not generalize well.

\subsection{Entire Space Adaptation Model (ESAM)}
To improve long-tail performance by suppressing inconsistent distribution between displayed and non-displayed items, ESAM utilizes discriminative domain adaptation with non-displayed items to extract better item features in the entire space. We mainly focus on the improvement on the item side $f_d$. Specifically, as shown in Fig.~\ref{figv31}, ESAM additionally introduces an unlabeled non-displayed item input stream with three constraints on $f_d$ of BaseModel.

\noindent\textbf{$\bullet$ Domain Adaptation with Attribute Correlation Alignment.}

Due to the existence of the domain shift between displayed items and non-displayed items. Hence, we utilize the DA technology to improve retrieval quality in the entire space.
\begin{figure}[t]
  \centering
  \includegraphics[width=0.95\linewidth]{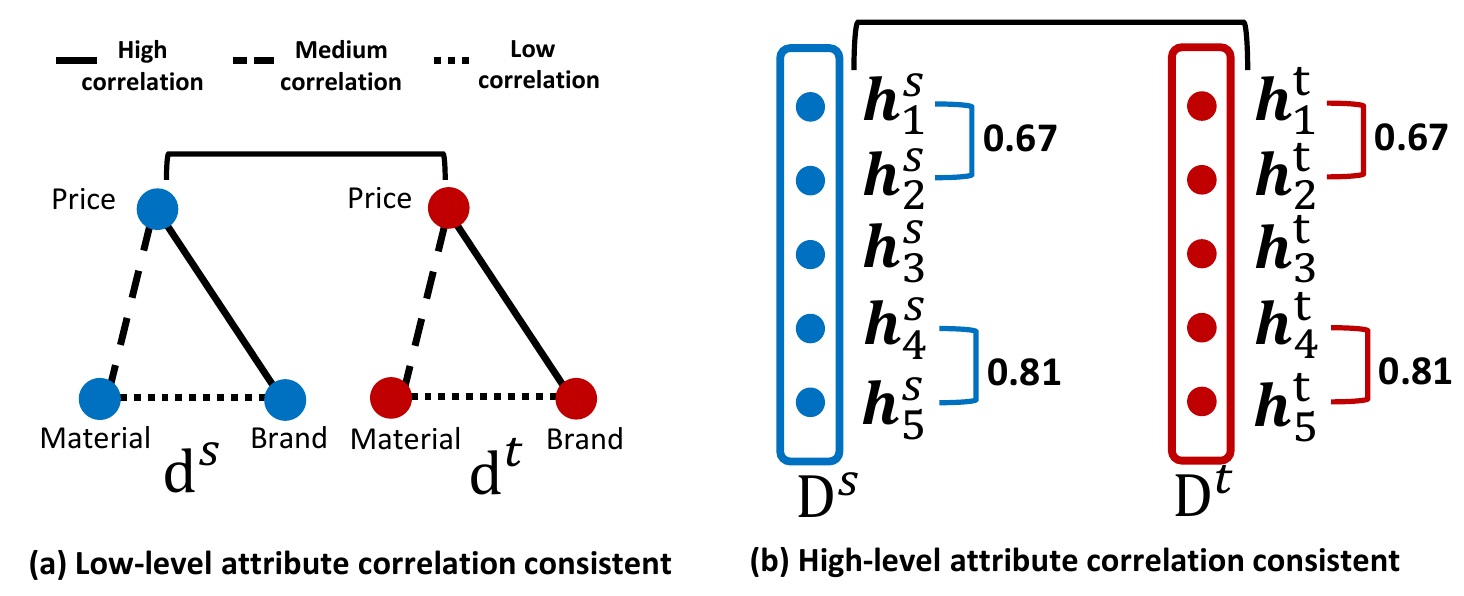}
  \caption{The correlation between low-level item attributes in the source domain is consistent with the target domain. Based on such constraint, we believe that the distributions of source and target domains are aligned when the correlation between high-level attribute vectors of item matrix in the source domain is consistent with the target domain.}
  \label{fig:3}
\end{figure}
Let $\bm{D}^s=[\bm{v}_{d_1^s}; \bm{v}_{d_2^s};...;\bm{v}_{d_{n}^s}] \in \mathbb{R}^{n \times L}$ represents the source item feature matrix for $\mathcal{D}^s_q$ and $\bm{D}^t=[\bm{v}_{d_1^t}; \bm{v}_{d_2^t};...;\bm{v}_{d_{n}^t}] \in \mathbb{R}^{n \times L}$  represents the target item feature matrix for $\mathcal{D}^t_q$, where $\bm{v}_{d_i^s} \in \mathbb{R}^{1 \times L}$ and $\bm{v}_{d_i^t} \in \mathbb{R}^{1 \times L}$ are source item feature and target item feature respectively and are produced by item side $f_d$. From another perspective, these two matrices can also be seen as a source high-level attribute matrix $\bm{D}^s=[\bm{h}_{1}^s, \bm{h}_{2}^s,...,\bm{h}_{L}^s] \in \mathbb{R}^{n \times L}$ and a target high-level attribute matrix $\bm{D}^t=[\bm{h}_{1}^t, \bm{h}_{2}^s,...,\bm{h}_{L}^t] \in \mathbb{R}^{n \times L}$, where $L$ is the dimension of the item feature, $\bm{h}_{j}^s \in \mathbb{R}^{n \times 1}$ and $\bm{h}_{j}^t \in \mathbb{R}^{n \times 1}$ represent the j-th high-level attribute vector in the source item matrix and the target item matrix, respectively.

We believe that the correlation between low-level attributes (\eg price, brand) of displayed items is consistent with non-displayed items, \eg as shown in Fig.~\ref{fig:3}, in the e-commerce, the brand attribute has a high correlation with the price attribute while the correlation between the brand and material attributes is low. This knowledge is consistent in the source and target domains. Therefore, we suggest that high-level attributes should also have the same correlation consistency as low-level attributes. Hence, we define the correlation matrix between high-level attribute vectors in the item matrix as the distribution. Specifically, we propose the attribute correlation congruence (A2C) as the distribution metric and reduce the distance between distributions to ensure that the correlation between high-level attributes is consistent in the source and target domains. Formally, the A2C is formulated as follows:
\begin{equation}\label{coral}
\begin{split}
L_{DA}&=\frac{1}{L^2}\sum_{(j,k)}(\bm{h}_{j}^{s\top}\bm{h}_{k}^{s}-\bm{h}_{j}^{t\top}\bm{h}_{k}^{t})^2\\
&=\frac{1}{L^2}\|Cov(\bm{D}^s)-Cov(\bm{D}^t)\|_F^2,\\
\end{split}
\end{equation}
where $\|\cdot\|_F^2$ denotes squared matrix Frobenius norm. $Cov(\bm{D}^s)\in \mathbb{R}^{L \times L}$ and $Cov(\bm{D}^t)\in \mathbb{R}^{L \times L}$ represent the covariance matrices of high-level item attributes, which can be computed as $Cov(\bm{D}^s)=\bm{D}^{s\top}\bm{D}^s$, and $Cov(\bm{D}^t)=\bm{D}^{t\top}\bm{D}^t$. The value at entry $(j,k)$ in the covariance matrix represents the correlation $\bm{h}_{j}^{\top}\bm{h}_{k}$ between $\bm{h}_{j}$ and $\bm{h}_{k}$ as shown in Fig.~\ref{fig:3}.

We take two batches of samples from source domain, and two batches of samples from target domain, and calculate $L_{DA}$ between each two covariance matrices extracted by BaseModel.  As shown in Table \ref{tab:lda}, the $L_{DA}$ between source and target domains is much larger than that between the same domains, which shows that there indeed exists the distribution shift.
\begin{table}
  \caption{Analysis of distribution distance ($L_{DA}$) extracted by BaseModel on the Industrial dataSet.}
  \label{tab:lda}
  \begin{tabular}{ccc}
    \toprule
    \textbf{Source-Source}&\textbf{Target-Target}&\textbf{Source-Target}\\
    \midrule
    9e-4 &8e-4&0.076\\
  \bottomrule
\end{tabular}
\end{table}

\noindent\textbf{$\bullet$ Center-Wise Clustering for Source Clustering.}

Most of BaseModel only optimize $L_s$, which is not insensitive to spatial structure in the feature space \cite{zhu2019improving} and make the model cannot learn discriminative features. In our preliminary study, we observe that for a query, items with same feedback are similar, and items with different feedbacks are dissimilar. Inspired by \cite{chen2019joint},  we propose a center-wise clustering to encourage the features of the items with the same feedback to be close together, and the features of the items with different feedbacks to move away from each other. The hinge-based center-wise clustering for each query $q$ can be formulated as:
\begin{align}
L_{DC}^{c} &=\sum_{j=1}^{n}\max(0,\Vert \frac{\bm{v}_{d_j^s}}{\Vert\bm{v}_{d_j^s}\Vert}-\mathbf{c}_q^{y_j^s}\Vert_2^2-m_1) \nonumber\\
&+\sum_{k=1}^{n_y}\sum_{u=k+1}^{n_y}\max(0,m_2-\Vert\mathbf{c}_q^k-\mathbf{c}_q^u\Vert_2^2)\label{cw} \\
\mathbf{c}_{q}^k &=\frac{\sum_{j=1}^n(\delta(y_j^s=Y_k)\cdot\frac{\bm{v}_{d_j^s}}{\Vert \bm{v}_{d_j^s}\Vert})}{\sum_{j=1}^n\delta(y_j^s=Y_k)}
\end{align}
where $n_y$ indicates the number of types of feedback (e.g. non-click, click, purchase and so on), and $y_j^s$ indicates the type of feedback on $d_j^s$. $\mathbf{c}^k_{q}$ is the class center of the features whose item take the same feedback $Y_k$ by $q$. And $m_1$ and $m_2$ are two distance constraint margins. $\delta(condition)=1$ if condition is satisfied.

In the right part of Eq.~\ref{cw}, the first term enforces intra-class (the items which have the same feedback) compactness, while the second term in Eq. \ref{cw} enforces the inter-class (the items which have different feedbacks) separability. Due to the existence of $L_{DA}$, source and target domains are highly-related. Therefore, it is reasonable to make the source item feature space more discriminative, such that the target item feature space will become discriminative also by optimizing Eq.~\ref{coral}. In this sense, the ranking model can extract better item features to improve domain adaptation, leading to better long-tail performance.

\noindent\textbf{$\bullet$ Self-Training for Target Clustering.}

For a query, we can assign a target pseudo-label (\ie positive or negative sample) for each non-displayed item. Right now, ESAM can be considered as a class-agnostic DA method that ignores target label information when aligning and may map the target items to the wrong positions (\eg matching target positive samples to source negative samples). The target samples with pseudo-labels will provide the model with target-discriminative information when aligning. Therefore, we use samples with pseudo-labels for self-training to increase the number of records of non-displayed items and mitigate negative transfer.

\begin{figure}[h]
  \centering
  \includegraphics[width=0.90\linewidth]{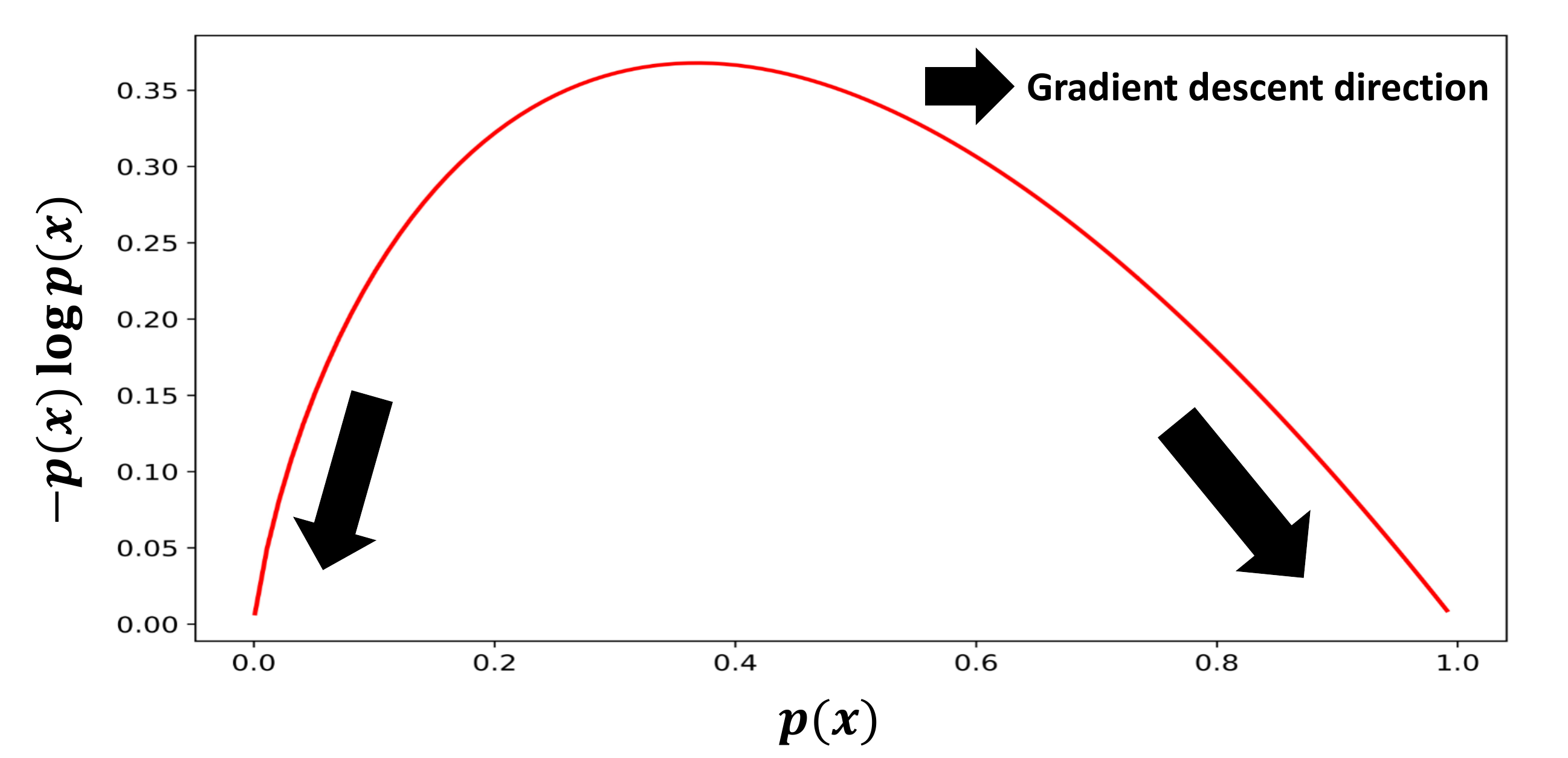}
  \caption{Entropy regularization $-p(x)\log p(x)$, which can be seen as a way to adopt self-training with pseudo-labels.}
  \label{fig:4}
\end{figure}

Specifically, minimizing the entropy regularization $-p \log p$ favors a low-density separation between classes and increase the target discrimination \cite{grandvalet2005semi}. We calculate the relevance score between query and non-displayed item as Eq. \ref{score} and convert it to $[0, 1]$ (e.g. the sigmoid operation). This score can be taken as the probability that the target item is a positive sample for this query. For click feedback, the score represents the probability that the query clicks the item. As shown in Fig.~\ref{fig:4}, optimizing entropy regularization by gradient descent enforces a score less than $0.4$ to gradually approach $0$ (negative sample), and a score greater than $0.4$ gradually approaches $1$ (positive sample). Therefore, this regularization can be seen as a way of self-training which increases the discriminative power between non-displayed items. However, when the regularization used directly, due to the large domain shift between displayed and non-displayed items in early training, the ranking model cannot predict non-displayed items correctly, especially for items whose score is in $[0.4, 0.6]$. Hence, the target samples are easily assigned false labels and stuck into the wrong feedback.  To this end, we adopt entropy regularization with a constraint to select reliable target samples for self-training:

\begin{footnotesize}
\begin{equation}\label{p}
L_{DC}^{p}=-\frac{\sum_{j=1}^{n}\delta(Sc_{{q,d_j^t}}<p_1|Sc_{{q,d_j^t}}>p_2)Sc_{{q,d_j^t}}\log Sc_{{q,d_j^t}}}{\sum_{j=1}^{n}\delta(Sc_{{q,d_j^t}}<p_1|Sc_{{q,j}}^t>p_2)},
\end{equation}
\end{footnotesize}

\noindent where $p_1$ and $p_2$ are two confidence thresholds that are used to select reliable samples with pseudo-labels of high confidence. $Sc_{{q,d_j^t}}$ is the relevance score between $q$ and $d_j^t$, which is calculated as Eq.~\ref{score} and converted to [0, 1]. $\delta(a|b)=1$ if condition $a$ or condition $b$ is satisfied. By optimizing Eq. \ref{p}, the model is self-trained with negative target items whose $Sc_{{q,d_j^t}}$ are less than $p_1$ and positive target items whose $Sc_{{q,d_j^t}}$  are greater than $p_2$.

This entropy regularization with a constraint guarantees the correctness of the target label to avoid the negative transfer. Moreover, based on curriculum learning \cite{bengio2009curriculum}, the model may learn target-discriminative information from reliable samples, thereby transforming more samples into reliable samples.
\subsection{Optimization}
In summary, the total loss of ESAM can be formulated as:
\begin{equation}
\begin{split}
L_{all}=L_{s}+\lambda_1L_{DA}+\lambda_2L_{DC}^{c}+\lambda_3L_{DC}^{p},
\end{split}
\end{equation}
where $\lambda_1$, $\lambda_2$ and $\lambda_3$ are hyperparameters which control the impact of corresponding terms. We define $\bm\Theta$ as the learnable weights of ESAM, which can be updated end-to-end by back propagation:
\begin{equation}
\bm\Theta^{t+1}=\bm\Theta^t-\eta\dfrac{\partial(L_{s}+\lambda_1L_{DA}+\lambda_2L_{DC}^{c}+\lambda_3L_{DC}^{p})}{\partial \bm \Theta},
\end{equation}
where $\eta$ is the learning rate. 
\section{applications}\label{application}
In this section, we apply our ESAM to two specific ranking scenarios: (i) item recommendation, and (ii) personalized search system.

\subsection{Item Recommendation}
Item recommendation is a typical ranking problem. One of the most important method for recommender systems is collaborative filtering which explores underlying user-item similarity~\cite{koren2009matrix}. The recommender system mainly includes four components, including feature composition ($q$ and $d$), feature extractor ($f_q$ and $f_d$), scoring function ($f_s$), and loss function ($L_s$).

\noindent \textbf{Feature Composition.} We mainly adopt three feature groups in item recommendation: user profile, user's behavior, and item profile. Each group is composed of some sparse features: user profile contains the user id, age, gender, etc; user's behavior contains the item ids that the user interacted recently; item profile contains the item id, brand, category, etc. Which feature groups and sparse features are used by the model depends on the design of the feature extractor.

\noindent \textbf{Feature Extractor.} The basic feature extractor of neural network-based recommendation systems is MLP  \cite{huang2013learning}. Currently, a variety of novel modules are designed to enhance the feature extraction, such as attention mechanism \cite{zhou2019deep},  transformer module \cite{chen2019behavior}.  In our experiments, we integrate ESAM into several extractors to prove that it is a general framework to improve long-tail performance.

\noindent \textbf{Scoring Function.} We define the scoring function as \cite{wang2017irgan}:
\begin{equation}\label{s1}
\begin{split}
Sc_{q,d}=f_s(\bm{v}_q,\bm{v}_d)=sigmoid(b_d+\bm{v}_q\bm{v}_d^{\top}),
\end{split}
\end{equation}
where $b_d$ is the bias term for item $d$, $\bm{v}_q,\bm{v}_d \in \mathbb{R}^{1\times L}$ are the features of user $q$ and item $d$ extracted from $f_q$ and $f_d$ respectively.

\noindent \textbf{Loss Function.} We treat click as feedback. Hence, for a query $q$, whether an item $d$ will be clicked can be regarded as a binary classification problem. We apply the point-wise cross-entropy objective function:
\begin{equation}
\begin{split}
L_{s}=-\frac{1}{n}\sum_{i=1}^{n}(y_{q,d_i}\log Sc_{q,d_i})+(1-y_{q,d_i})\log (1-Sc_{q,d_i}),
\label{auc}
\end{split}
\end{equation}
where $y_{q,d_i}$ is a binary variable indicating whether the query clicked on the item $d_i$ ($y_{q,d_i}=1$ if query $q$ clicked on item $d_i$, and $y_{q,d_i}=0$ otherwise), $n$ is the number of displayed items for query $q$.

The above model is trained using only $L_s$, which considers only the displayed items. As shown in Sec.~\ref{method} and Fig.~\ref{figv31}, to mitigate domain shift for performance improvement of model, we randomly assign non-displayed items to each query in every training epoch and integrate ESAM into the model by introducing additional non-display item input stream with three constraints on item side $f_d$.

\subsection{Personalized Search System}
The personalized search system investigated in this work is to identify all possible items at which the user could be interested based on the keyword issued by the user. Hence, the integration of ESAM, feature extractor, and loss function are the same as item recommendation, and only the feature composition and scoring function are different. For the feature composition, personalized search systems introduce an additional keyword feature group. The keyword is transformed into a vector by averaging word vectors of the keyword. For the scoring function, we define the cosine similarity as scoring function \cite{wang2017irgan}:
\begin{equation}\label{consine}
\begin{split}
Sc_{q,d}=f_s(\bm{v}_q,\bm{v}_d)=\frac{\bm{v}_q\bm{v}_d^{\top}}{\|\bm{v}_q\|\cdot\|\bm{v}_d\|}.
\end{split}
\end{equation}
\section{experiments}\label{experiments}
In this section, we conduct our experiments corresponding to two real-world applications of ESAM as discussed above, i.e., item recommendation and personalized search system.
\subsection{Datasets}

\noindent\textbf{$\bullet$ Item Recommendation.}

\textbf{MovieLens-1M \footnote{http://www.grouplens.org/datasets/movielens/}:} The dataset consists of 1 million movie rating records over thousands of movies and users. We first binarize ratings to simulate the CTR prediction task, i.e., the ratings greater than 3 are displayed positive samples, and the rest are displayed negative samples, which is common to this problem setting \cite{volkovs2015effective}. The features we used are the user id, age, gender, occupation of the user, user behavior in query side, and movie id, release year, genres of the movie in item side. At each training epoch, we randomly add $10$ non-displayed movies similar to the displayed movies (\ie $n=10$). The similar movies are the non-displayed movies whose genres are the same as displayed movies. We divide data randomly into the training, validation, and test sets by a ratio of 8: 1: 1.

\noindent\textbf{$\bullet$ Personalized Search System.}

\textbf{CIKM Cup 2016 Personalized E-Commerce Search\footnote{https://competitions.codalab.org/competitions/11161/}:} The Personalized E-commerce Search Challenge releases a dataset provided by DIGINETICA. Each item contains item id, category id, title, and description, each query consists of the user id, user's gender, age, history behavior, and  keyword id (transformed by a list of words) issued by the user, and  each record contains a query, 10 displayed items for this query and the items that the query clicked. In addition, we adopt the same strategy as MovieLens-1M to assign the non-displayed items for each record. For training, validation and test sets construction, we apply a 8:1:1 random splitting.

\textbf{Industrial Dataset of Taobao:} We collect the logs of Taobao for one week in June 2019. In this dataset, all the features used, the way of obtaining non-displayed items, and the organization of the records are the same as CIKM Cup 2016 dataset. In the experiments, we split the dataset into three parts: training set (records from the first day to the fifth day), validation set (records from the sixth day) and test set (records from the seventh day). The basic statistics of these three datasets are reported in Table \ref{tab:data}.
\begin{table}
  \caption{Statistics of experimental datasets. 'M' represents the million and 'B' represents the billion.}
  \label{tab:data}
  \begin{tabular}{cccc}
    \toprule
    \textbf{Dataset}&\textbf{\#Query}&\textbf{\#Item}&\textbf{\#Record}\\
    \midrule
    MovieLens-1M dataset & 6040&3883&1M\\
    CIKM Cup 2016 dataset & 5.7M&13.4M&24.5M\\
    Taobao Industrial dataset &114.6M&223.2M&11.9B\\
  \bottomrule
\end{tabular}
\end{table}
\begin{table*}[h]
  \centering
  \caption{Performance comparison between the methods without and with ESAM on MovieLens-1M dataset. ``Hot'' represents hot items in the test set, ``Long-tail'' represents long-tail items in the test set, and ``Entire'' represents all items in the test set. The best results are highlighted in boldface. The improvements over the baseliens are statistically significance at $0.05$ level.}
  \label{tab:movie}\resizebox{\linewidth}{!}{
    \begin{tabular}{cccccccccccccccccccc}
   \toprule
    \multirow{3}{*}{Group}&\multirow{3}{*}{Method}&
    \multicolumn{9}{c}{w/o ESAM}&\multicolumn{9}{c}{w/ ESAM}\cr
    \cmidrule(lr){3-11} \cmidrule(lr){12-20}
    &&\multicolumn{3}{c}{NDCG@20}&\multicolumn{3}{c}{Recall@20}&\multicolumn{3}{c}{MAP}&\multicolumn{3}{c}{NDCG@20}&\multicolumn{3}{c}{Recall@20}&\multicolumn{3}{c}{MAP}\cr
    \cmidrule(lr){3-5} \cmidrule(lr){6-8}\cmidrule(lr){9-11} \cmidrule(lr){12-14} \cmidrule(lr){15-17} \cmidrule(lr){18-20}
    &&Hot&Long-tail&Entire&Hot&Long-tail&Entire&Hot&Long-tail&Entire&Hot&Long-tail&Entire&Hot&Long-tail&Entire&Hot&Long-tail&Entire\cr
    \midrule
    \multirow{4}{*}{Single Domain}&NeuralMF \cite{huang2013learning}&0.2296&0.0508&0.2017&0.2660&0.0732&0.2453&0.1953&0.0486&0.1813&\textbf{0.2431}&\textbf{0.0812}&\textbf{0.2238}&\textbf{0.2753}&\textbf{0.1005}&\textbf{0.2644}&\textbf{0.2014}&\textbf{0.0674}&\textbf{0.1962}\cr
    &YoutubeNet \cite{covington2016deep}&0.2519&0.0983&0.2385&0.2976&0.1142&0.2691&0.2138&0.0874&0.2057&\textbf{0.2671}&\textbf{0.1264}&\textbf{0.2613}&\textbf{0.3125}&\textbf{0.1369}&\textbf{0.2947}&\textbf{0.2282}&\textbf{0.1108}&\textbf{0.2270}\cr
    &RALM \cite{liu2019real}&0.2658&0.1237&0.2584&0.3081&0.1294&0.2883&0.2259&0.1037&0.2236&\textbf{0.2784}&\textbf{0.1493}&\textbf{0.2738}&\textbf{0.3249}&\textbf{0.1455}&\textbf{0.3057}&\textbf{0.2376}&\textbf{0.1271}&\textbf{0.2412}\cr
    &BST\cite{chen2019behavior}&0.2816&0.1371&0.2835&0.3292&0.1373&0.3104&0.2429&0.1230&0.2481&\textbf{0.2935}&\textbf{0.1627}&\textbf{0.2984}&\textbf{0.3453}&\textbf{0.1516}&\textbf{0.3295}&\textbf{0.2579}&\textbf{0.1451}&\textbf{0.2608}\cr
    \midrule
     \multirow{2}{*}{Domain Adaptation}&DALRF \cite{tran2019domain}&0.2731&0.1503&0.2467&0.3092&0.1424&0.2953&0.2376&0.1431&0.2359&\textbf{0.2816}&\textbf{0.1584}&\textbf{0.2503}&\textbf{0.3158}&\textbf{0.1507}&\textbf{0.3012}&\textbf{0.2431}&\textbf{0.1479}&\textbf{0.2418} \cr
    &DARec \cite{yuan2019darec}&0.2748&0.1565&0.2497&0.3124&0.1483&0.2996&0.2418&0.1488&0.2401&\textbf{0.2824}&\textbf{0.1631}&\textbf{0.2614}&\textbf{0.3182}&\textbf{0.1549}&\textbf{0.3079}&\textbf{0.2503}&\textbf{0.1579}&\textbf{0.2495} \cr
    \bottomrule
    \end{tabular}}
\end{table*}
\begin{table*}[h]
  \centering
  \caption{Performance comparison between methods without and with ESAM on CIKM Cup 2016 dataset. The best results are highlighted in boldface. The improvements over the baseliens are statistically significance at $0.05$ level.}
  \label{tab:CIKM}\resizebox{\linewidth}{!}{
    \begin{tabular}{cccccccccccccccccccc}
    \toprule
    \multirow{3}{*}{Group}&\multirow{3}{*}{Method}&
    \multicolumn{9}{c}{w/o ESAM}&\multicolumn{9}{c}{w/ ESAM}\cr
    \cmidrule(lr){3-11} \cmidrule(lr){12-20}
    &&\multicolumn{3}{c}{NDCG@20}&\multicolumn{3}{c}{Recall@20}&\multicolumn{3}{c}{MAP}&\multicolumn{3}{c}{NDCG@20}&\multicolumn{3}{c}{Recall@20}&\multicolumn{3}{c}{MAP}\cr
    \cmidrule(lr){3-5} \cmidrule(lr){6-8}\cmidrule(lr){9-11} \cmidrule(lr){12-14} \cmidrule(lr){15-17} \cmidrule(lr){18-20}
    &&Hot&Long-tail&Entire&Hot&Long-tail&Entire&Hot&Long-tail&Entire&Hot&Long-tail&Entire&Hot&Long-tail&Entire&Hot&Long-tail&Entire\cr
    \midrule
    \multirow{4}{*}{Single Domain}&NeuralMF \cite{huang2013learning}&0.1840&0.0521&0.1296&0.3145&0.0947&0.2265&0.1596&0.0559&0.1158&\textbf{0.2043}&\textbf{0.1085}&\textbf{0.1543}&\textbf{0.3420}&\textbf{0.1495}&\textbf{0.2652}&\textbf{0.1718}&\textbf{0.0857}&\textbf{0.1409}\cr
    &YoutubeNet \cite{covington2016deep}&0.2196&0.0942&0.1572&0.3481&0.1422&0.2594&0.1928&0.0974&0.1437&\textbf{0.2347}&\textbf{0.1295}&\textbf{0.1821}&\textbf{0.3702}&\textbf{0.1735}&\textbf{0.2859}&\textbf{0.2139}&\textbf{0.1266}&\textbf{0.1703}\cr
    &RALM \cite{liu2019real}&0.2341&0.1195&0.1794&0.3592&0.1583&0.2739&0.2164&0.1219&0.1682&\textbf{0.2457}&\textbf{0.1408}&\textbf{0.2049}&\textbf{0.3796}&\textbf{0.1841}&\textbf{0.2985}&\textbf{0.2295}&\textbf{0.1451}&\textbf{0.1879}\cr
    &BST\cite{chen2019behavior}&0.2483&0.1392&0.2067&0.3824&0.1793&0.3021&0.2319&0.1436&0.1928&\textbf{0.2607}&\textbf{0.1586}&\textbf{0.2241}&\textbf{0.3976}&\textbf{0.2058}&\textbf{0.3249}&\textbf{0.2465}&\textbf{0.1620}&\textbf{0.2194}\cr
    \midrule
    \multirow{2}{*}{Domain Adaptation}&DALRF \cite{tran2019domain}&0.2213&0.1456&0.1861&0.3684&0.1921&0.2973&0.2024&0.1583&0.1895&\textbf{0.2362}&\textbf{0.1578}&\textbf{0.2014}&\textbf{0.3805}&\textbf{0.2052}&\textbf{0.3097}&\textbf{0.2152}&\textbf{0.1735}&\textbf{0.2008} \cr
    &DARec \cite{yuan2019darec}&0.2259&0.1502&0.1923&0.3705&0.1982&0.3018&0.2079&0.1604&0.1936&\textbf{0.2348}&\textbf{0.1683}&\textbf{0.2091}&\textbf{0.3859}&\textbf{0.2053}&\textbf{0.3084}&\textbf{0.2137}&\textbf{0.1762}&\textbf{0.2041} \cr
    \bottomrule
    \end{tabular}}
\end{table*}
\begin{table}[h]
  \centering
  \caption{Cold-start performance of BST \cite{chen2019behavior} w/o and w/ ESAM.}
  \label{tab:cold}\resizebox{\linewidth}{!}{
    \begin{tabular}{ccccccc}
    \toprule
    \multirow{2}{*}{Dataset}&
    \multicolumn{3}{c}{w/o ESAM}&\multicolumn{3}{c}{w/ ESAM}\cr
    \cmidrule(lr){2-4} \cmidrule(lr){5-7}
    &NDCG@20&Recall@20&MAP&NDCG@20&Recall@20&MAP\cr
    \midrule
    MovieLens-1M&0.0319&0.0621&0.0485&\textbf{0.0937}&\textbf{0.1233}&\textbf{0.0949}\cr
    CIKM Cup 2016&0.0436&0.0759&0.0531&\textbf{0.0982}&\textbf{0.1349}&\textbf{0.1023}\cr
    \bottomrule
    \end{tabular}}
\end{table}
\subsection{Baselines and Evaluation Metrics}
To verify that ESAM is a general framework, we integrate it to some single-domain ranking approaches based on neural network. Note that, The feature extractor ($f_q$ and $f_d$) is consistent in personalized search and recommendation applications, only the feature composition ($q$ and $d$), and scoring function ($f_s$) depend on the application. The specific differences can be seen in Sec. \ref{application}. We also compare some DA-based ranking models and "missing not random" methods to show the superiority of our proposed ESAM.

Specifically, the single-domain methods are: NeuralMF \cite{huang2013learning}, RALM \cite{liu2019real}, YoutubeNet \cite{covington2016deep}, BST \cite{chen2019behavior}. The DA-based methods are: DARec \cite{yuan2019darec}, DA learning to rank framework (DALRF) \cite{tran2019domain}. A "missing not random" method is: unbiased imputation model (UIM)\cite{yuan2019improving}. For methods based on DA to improve long-tail performance (e.g., DALRF, DARec), we only replace the original DA constraint with ESAM, and keep other modules unchanged for comparison. Note that the "missing not random" method requires an unbiased dataset. Here the unbiased dataset collected through deploying a uniform policy on a small percentage of our online traffic. We only replace the constraints that use the unbiased dataset with ESAM for fair comparison. Three standard evaluation metrics, NDCG (Normalized Discounted Cumulative Gain), Recall and MAP (Mean Average Precision) are used for performance comparison.

\subsection{Implementation Details}
 All methods are implemented with TensorFlow and trained with Adam optimizer. All methods are trained five times and the average results are reported. For the hyperparameters $\lambda_1$, $\lambda_2$ and $\lambda_3$, we search the best ones from $\{0.01,0.1,0.3,0.5,0.7,1,10\}$ and set $\lambda_1=0.7$, $\lambda_2=0.3$ and $\lambda_3=0.5$. For other hyperparameters, we set learning rate $\eta=1e-4$, batch size to be $256$, feature dimension $L=128$, distance constraints $m_1=0.2$ and $m_2=0.7$, and confidence thresholds $p_1=0.2$ and $p_2=0.8$. For the hyperparameters of comparison methods, we use a grid search based on performance on the validation set. Note that, we assign each query $10$ source items $\mathcal{D}^s_q$ and 10 target items $\mathcal{D}^t_q$ (\ie $n=10$).

\subsection{Performance Comparison}
We divide the entire item space into hot and long-tail items by the display frequency. Through the inflection point shown in Fig.~\ref{figv21}b, we define the top 20\% items as hot items and the rest as long-tail items.

\textbf{Public Datasets: } Table \ref{tab:movie} and Table \ref{tab:CIKM} show the performance of baselines without and with ESAM on the Movielens-1M dataset and CIKM Cup 2016 dataset, respectively. We make the following observations from the results. (1) The performance of each model for hot items is always much higher than that of long-tail items, which indicates that the distributions of hot and long-tail items are inconsistent. (2) The single-domain baselines with ESAM outperform that without ESAM. For the personalized search application, the baselines with ESAM achieve the average absolute NDCG@20 / Recall@20 / MAP  gain of 1.4\% / 2.1\% / 1.5\% in the hot space,  3.3\% / 3.4\% / 2.5\% in the long-tail space and 2.3\% / 2.8\% /2.3\%  in the entire space; for the item recommendation application, the baselines with ESAM achieve the average absolute  NDCG@20 / Recall@20 / MAP gain of 1.3\% / 1.4\% / 1.2\% in the hot space,  2.7\% / 2.0\% / 2.2\% in the long-tail space and 1.8\% / 1.9\% / 1.7\% in the entire space. The significant improvement in the long-tail space proves that the proposed ESAM is effective to alleviate the domain shift problem for long-tail items, making baselines learn better features in the entire space for different datasets and applications. (3) We find that ESAM outperforms other DA methods (\ie DALRF, DARec). The results confirm that the discriminative domain adaptation we design, which considers the correlation between high-level attributes and adopts center-wise clustering and self-training to improve source discrimination and target discrimination, can effectively transfer more discriminative knowledge.
\begin{table}[h]
  \centering
  \caption{Ablation study on Industrial dataset.}
  \label{tab:abla}\resizebox{\linewidth}{!}{
    \begin{tabular}{ccccccc}
    \toprule
    \multirow{2}{*}{Method}&
    \multicolumn{3}{c}{Recall@1k}&\multicolumn{3}{c}{Recall@3k}\cr
    \cmidrule(lr){2-4} \cmidrule(lr){5-7}
    &Hot&Long-tail&Entire&Hot&Long-tail&Entire\cr
    \midrule
    BaseModel (BM)&0.1842&0.1067&0.1298&0.2776&0.1845&0.2149\cr
    BM+$L_{DA}$ &0.1921&0.1364&0.1453&0.2805&0.2162&0.2409\cr
    ESAM w/o $L_{DA}$ &0.1976&0.1298&0.1439&0.2873&0.2043&0.2318\cr
    ESAM w/o $L_{DC}^c$&0.1926&0.1392&0.1507&0.2812&0.2195&0.2490\cr
    ESAM w/o $L_{DC}^p$&0.1975&0.1413&0.1532&0.2886&0.2207&0.2504\cr
    ESAM&\textbf{0.1982}&\textbf{0.1567}&\textbf{0.1637}&\textbf{0.2894}&\textbf{0.2383}&\textbf{0.2621}\cr
    \midrule
    UIM\cite{yuan2019improving}&0.1892&0.1456&0.1543&0.2831&0.2349&0.2675\cr
    UIM$_{ESAM}$&\textbf{0.1931}&\textbf{0.1521}&\textbf{0.1608}&\textbf{0.2905}&\textbf{0.2417}&\textbf{0.2759}\cr
    \bottomrule
    \end{tabular}}
\end{table}

\textbf{Cold-Start Performance: }We randomly select 20\% of the records in the test set and delete from the training set all records whose displayed items containing these selected test records. We employ BST \cite{chen2019behavior} as the BaseModel since it has the best performance on the public dataset. As shown in Table \ref{tab:cold}, we find that BST is difficult to solve the cold-start problem due to the lack of these items in the displayed space. While ESAM can significantly improve the cold-start performance because it introduces non-displayed items which contain the cold-start items, to enhance the feature learning.
\begin{figure*}[ht]\label{tsne}
    \centering
  \begin{subfigure}[b]{0.22\textwidth}
    \includegraphics[width=\textwidth]{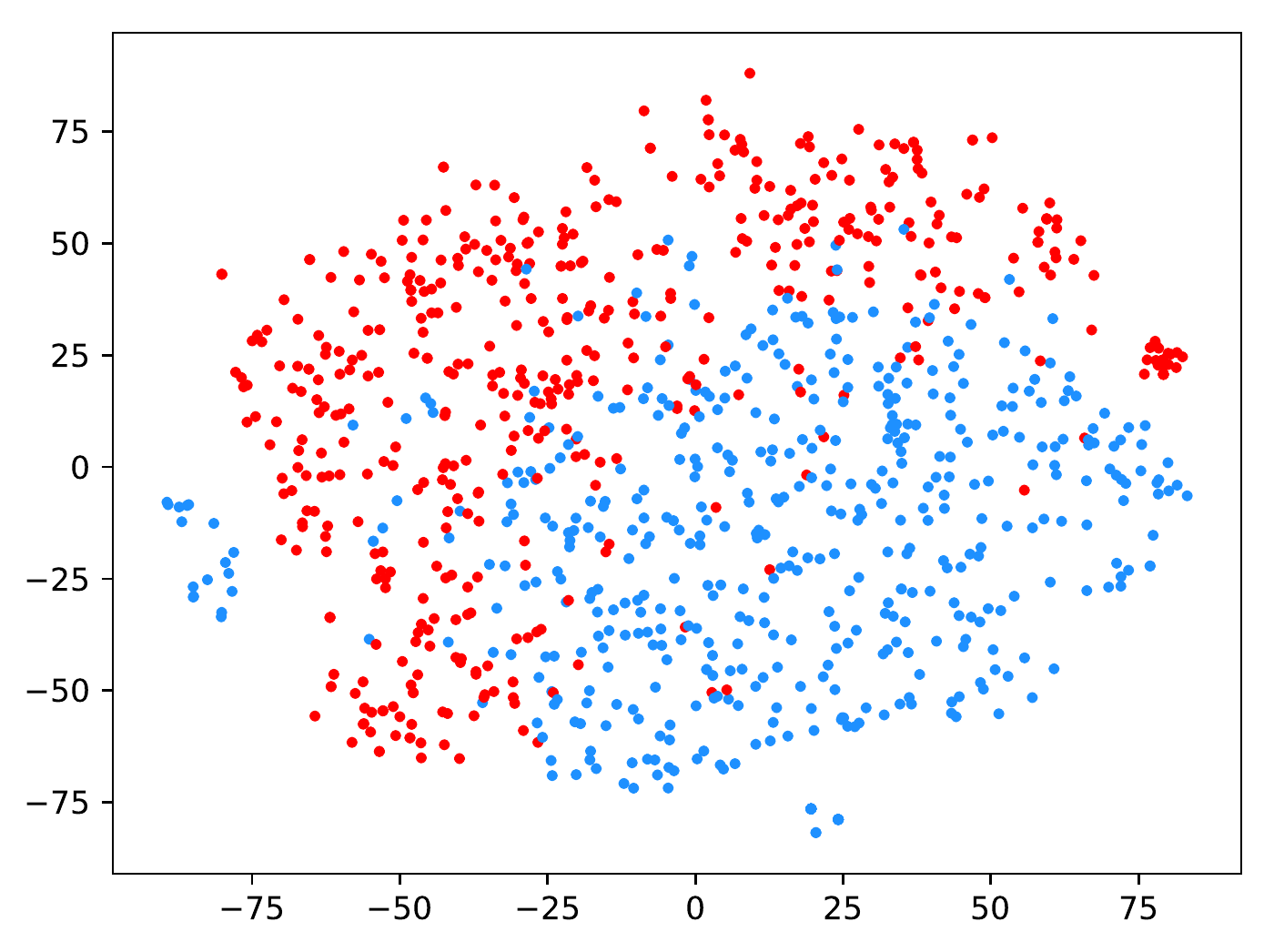}
    \caption{$L_s$}
    \label{2D1}
  \end{subfigure}
   \begin{subfigure}[b]{0.22\textwidth}
    \includegraphics[width=\textwidth]{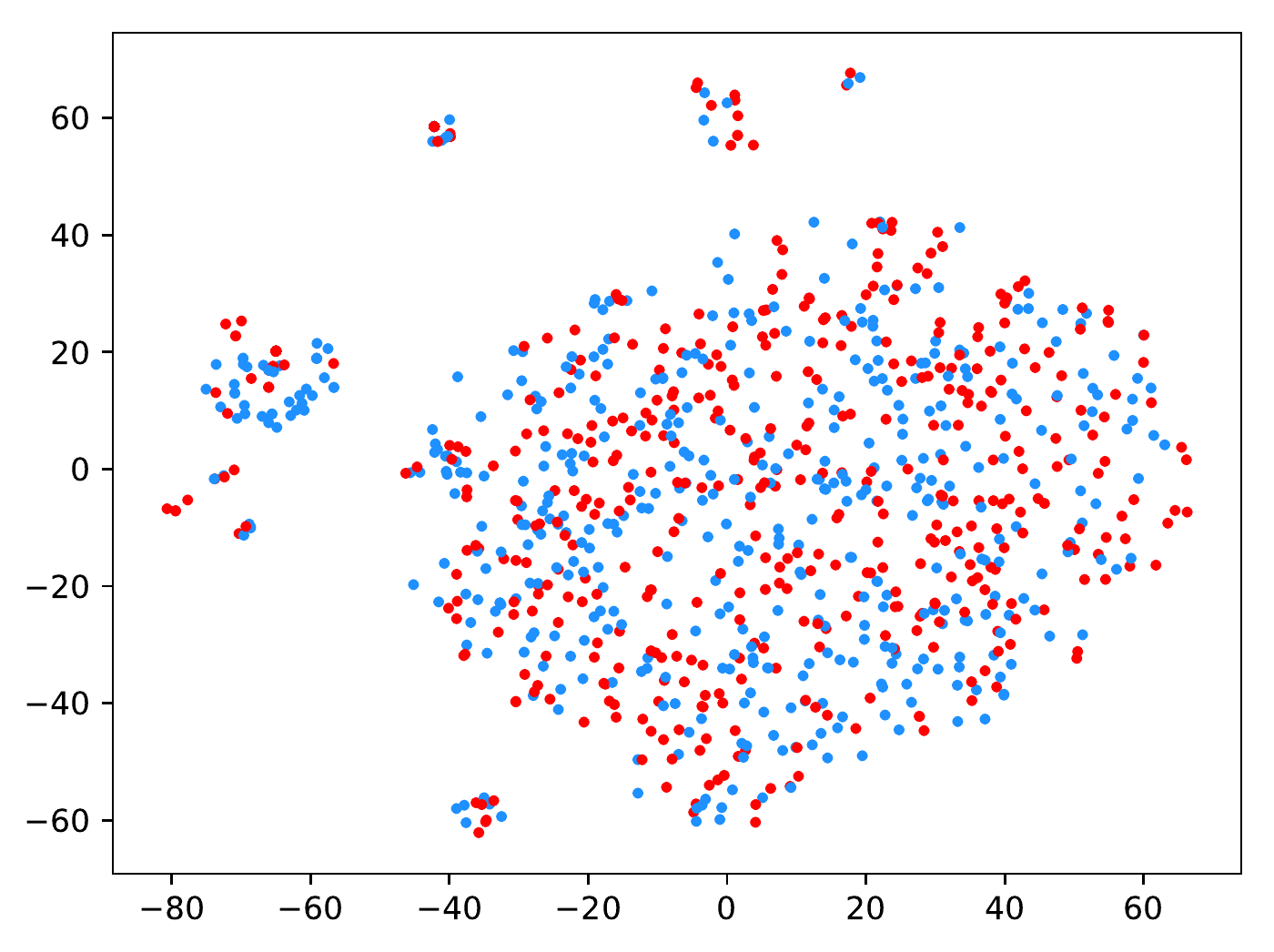}
    \caption{$L_{s}+L_{DA}$}
    \label{2D2}
  \end{subfigure}
    \begin{subfigure}[b]{0.22\textwidth}
    \includegraphics[width=\textwidth]{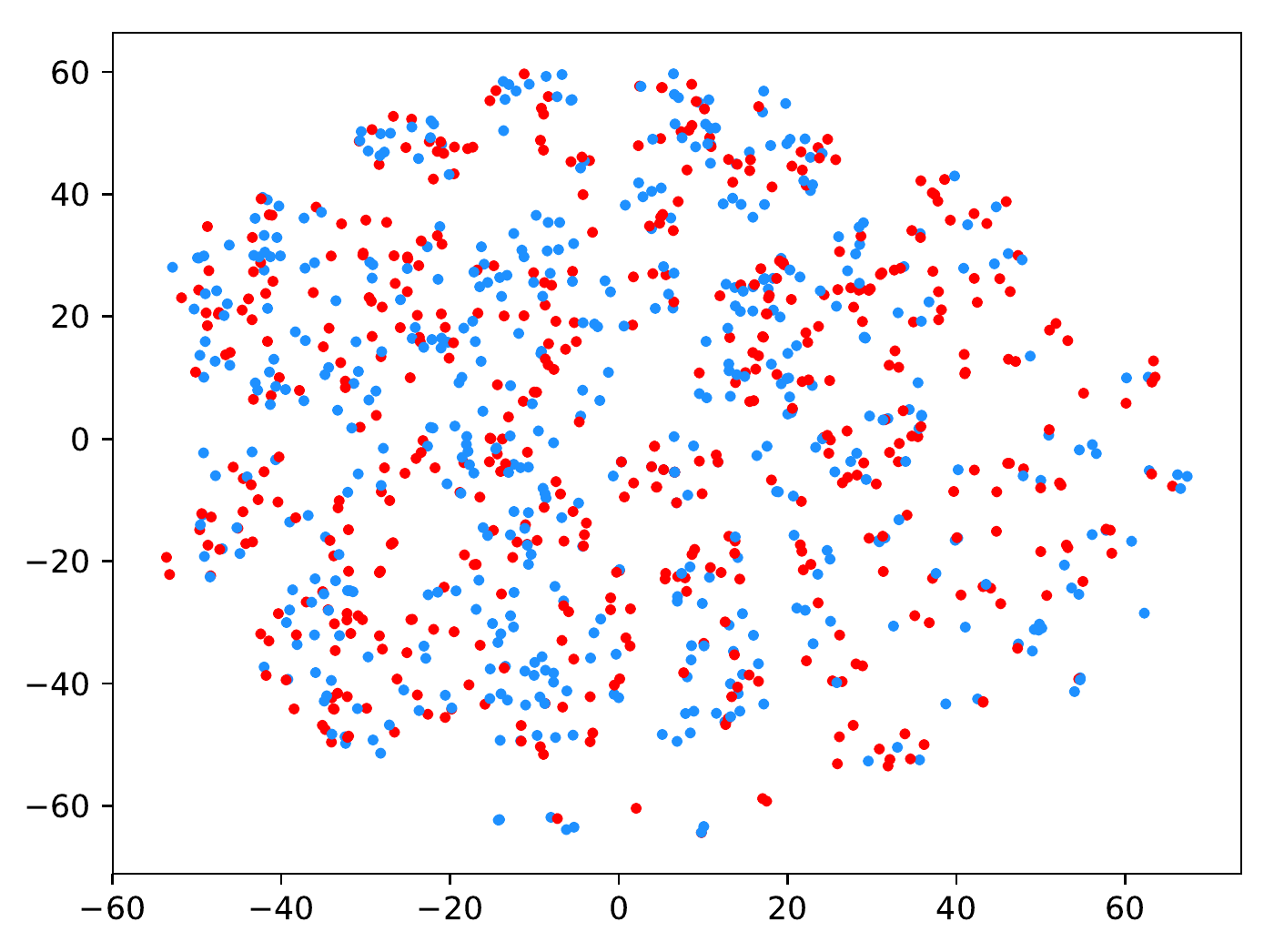}
    \caption{$L_{s}+L_{DA}+L_{DC}^c$}
    \label{2D3}
  \end{subfigure}
    \begin{subfigure}[b]{0.22\textwidth}
    \includegraphics[width=\textwidth]{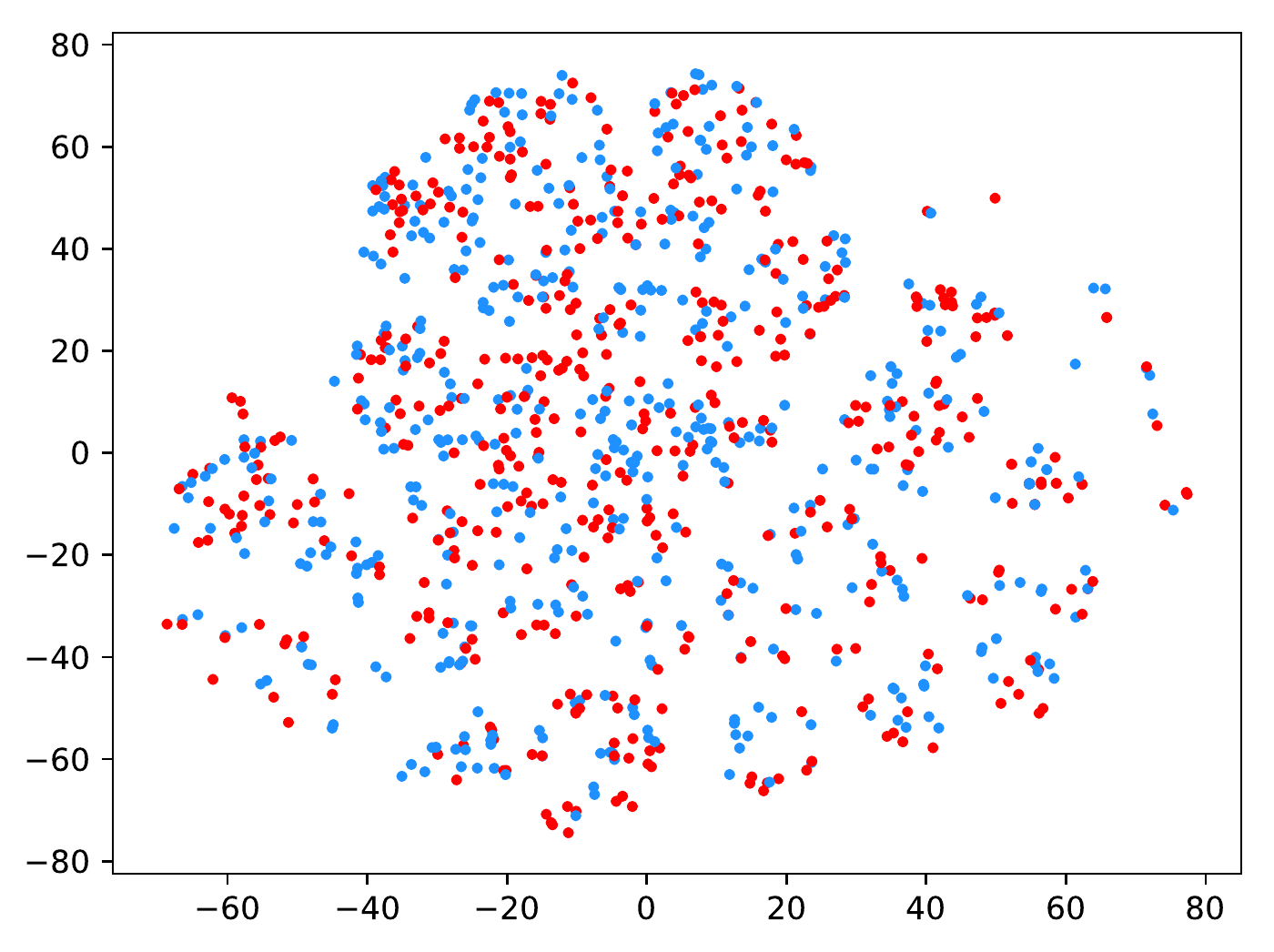}
    \caption{$L_{s}+L_{DA}+L_{DC}^c+L_{DC}^p$}
    \label{2D4}
  \end{subfigure}
\caption{T-SNE visualization on Industrial dataset. Red and blue points represent target and source samples respectively.}
\label{tsne}
\end{figure*}
\begin{figure*}[ht]
    \centering
  \begin{subfigure}[b]{0.22\textwidth}
    \includegraphics[width=\textwidth]{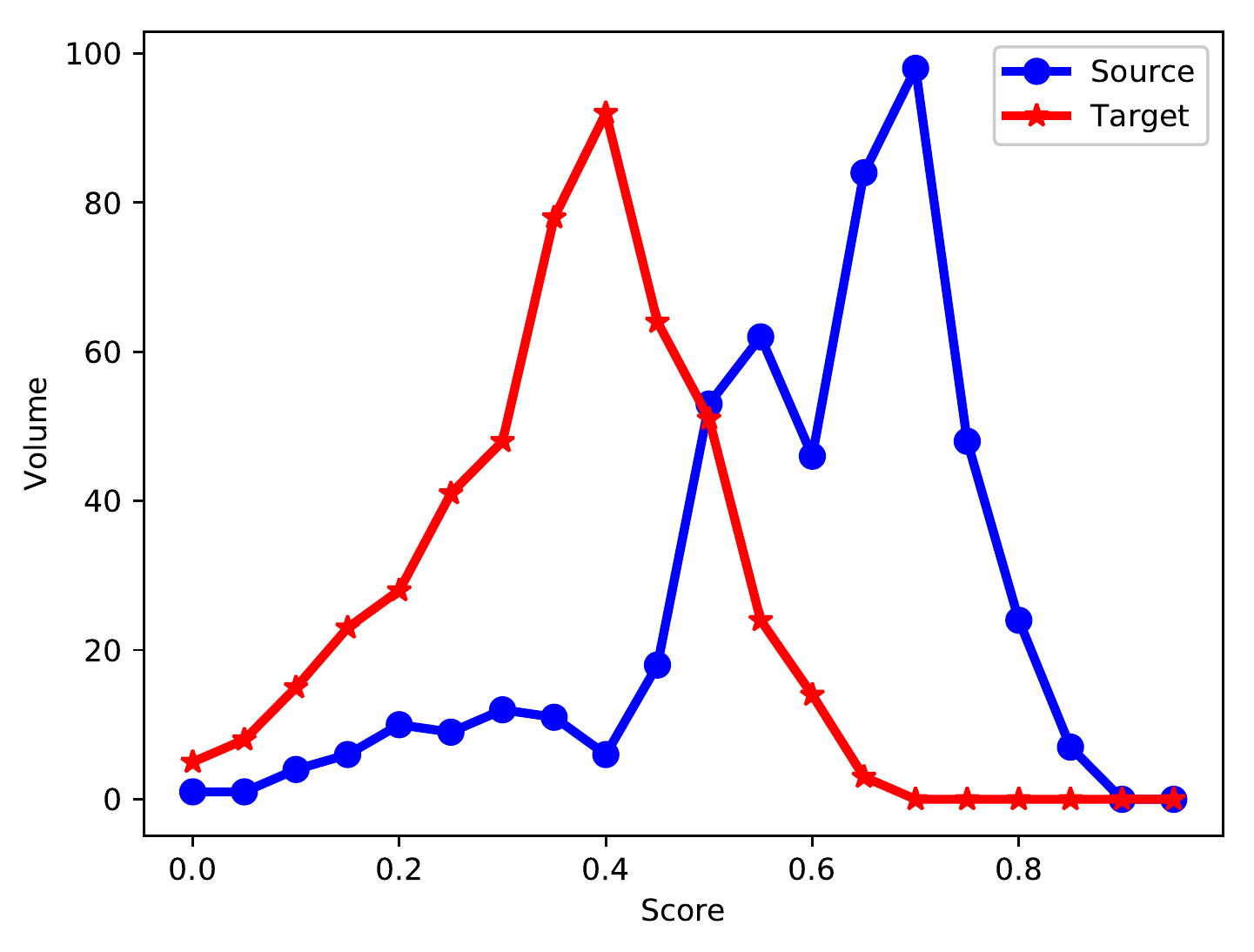}
    \caption{$L_s$}
    \label{s2D1}
  \end{subfigure}
   \begin{subfigure}[b]{0.22\textwidth}
    \includegraphics[width=\textwidth]{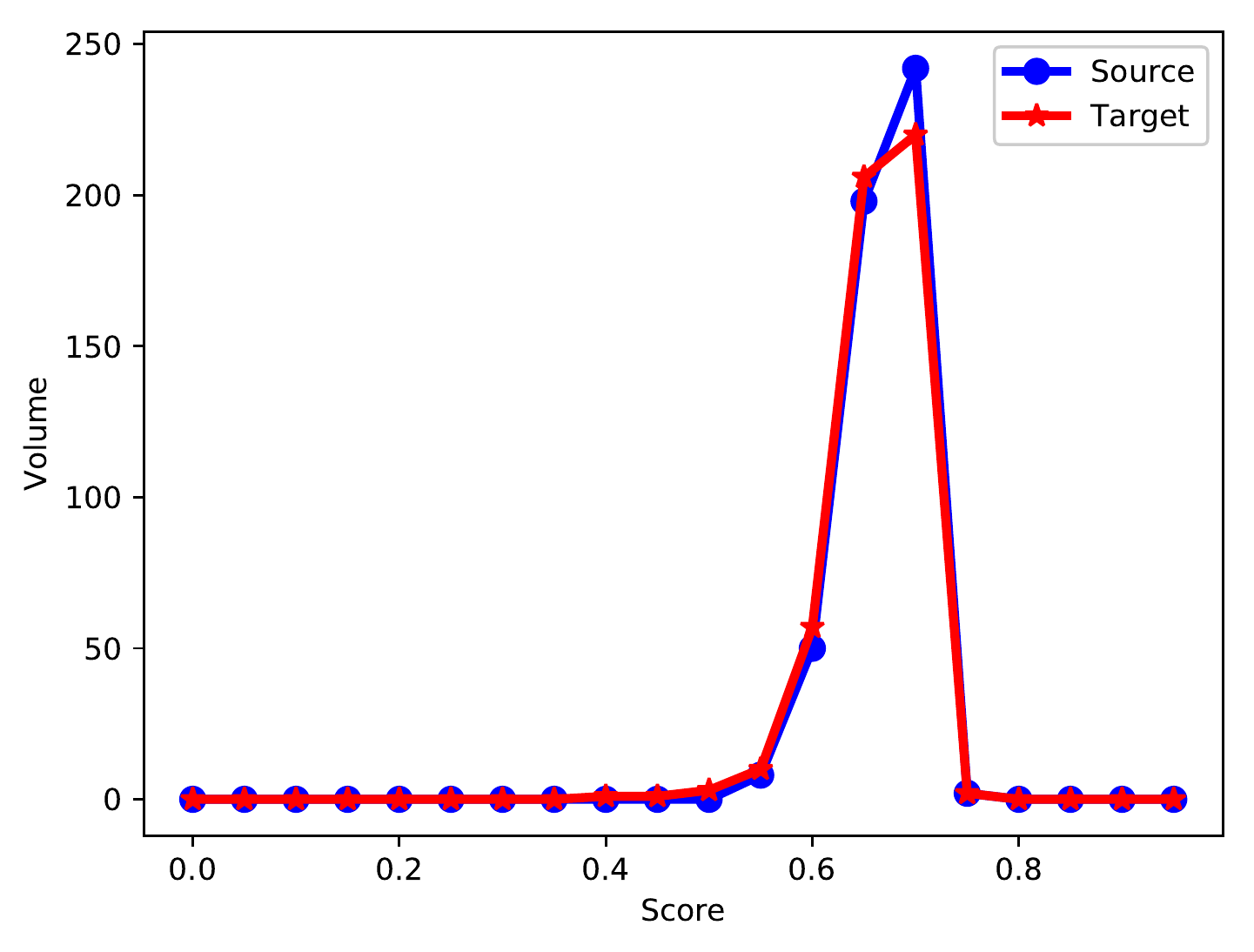}
    \caption{$L_{s}+L_{DA}$}
    \label{s2D2}
  \end{subfigure}
    \begin{subfigure}[b]{0.22\textwidth}
    \includegraphics[width=\textwidth]{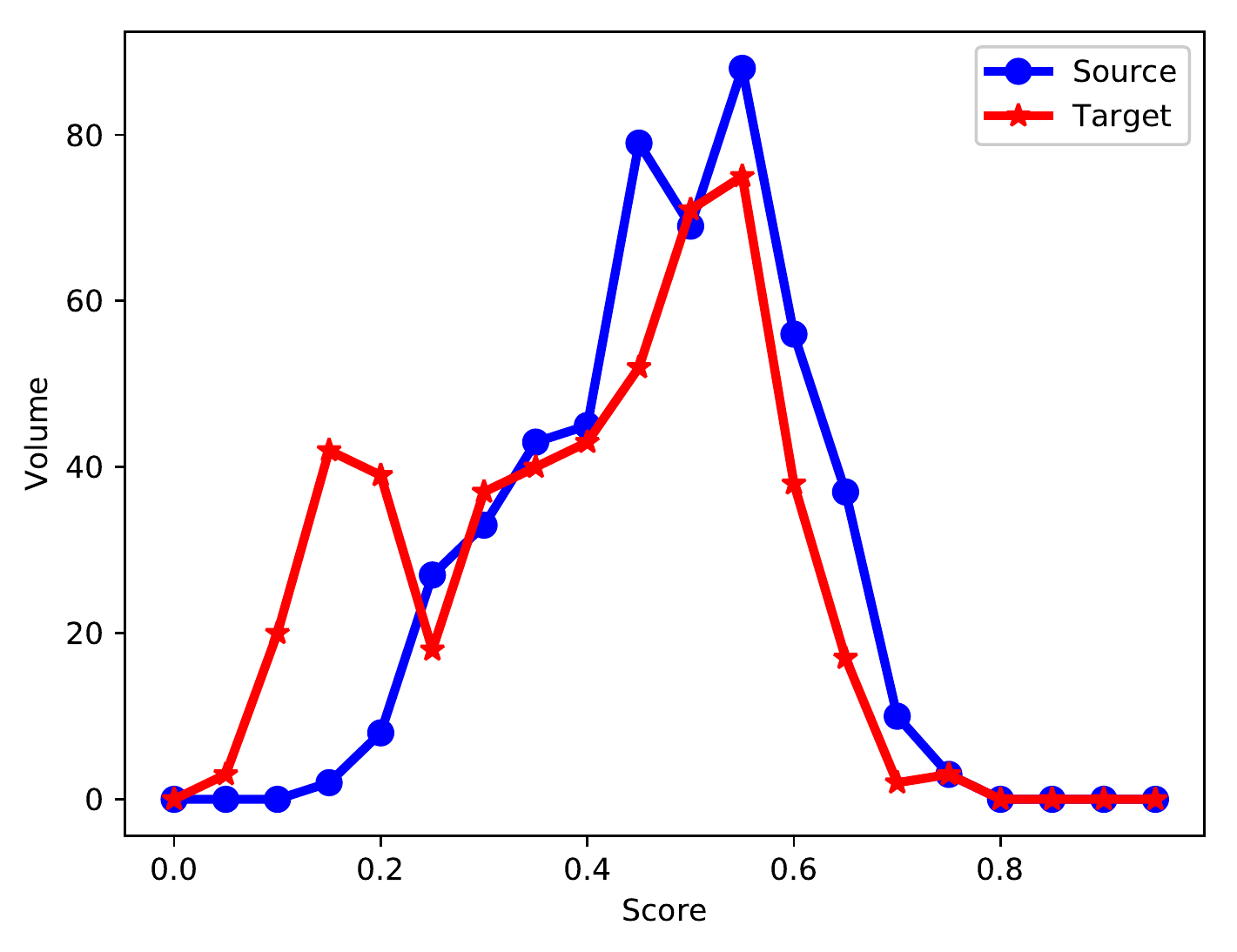}
    \caption{$L_{s}+L_{DA}+L_{DC}^c$}
    \label{s2D3}
  \end{subfigure}
    \begin{subfigure}[b]{0.22\textwidth}
    \includegraphics[width=\textwidth]{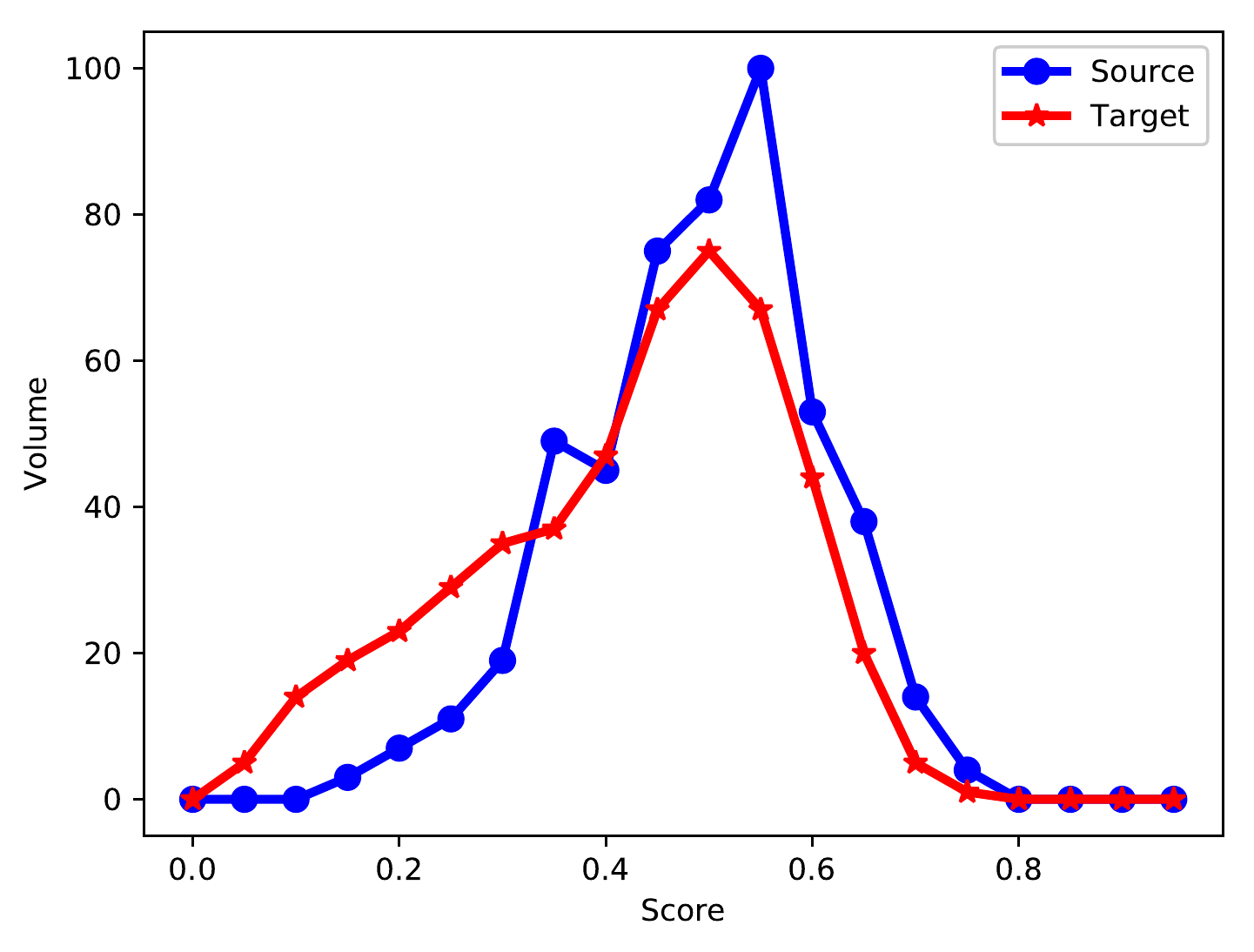}
    \caption{$L_{s}+L_{DA}+L_{DC}^c+L_{DC}^p$}
    \label{s2D4}
  \end{subfigure}
\caption{Score distribution on Industrial dataset. Red and blue represent the target and source score distributions respectively.}
\label{scoref}
\end{figure*}

\textbf{Industrial Dataset of Taobao:} Because the Taobao Industrial dataset is very large, we adopt Recall@1k and Recall@3k as the metrics and use the BST \cite{chen2019behavior} that has the best performance on the public dataset as the BaseModel for ablation study. Several observations can be made based on Table \ref{tab:abla}. (1) Comparing the method w/o $L_{DA}$ and that w/ $L_{DA}$ (\eg BM and BM+$L_{DA}$ or ESAM w/o $L_{DA}$ and ESAM), we find that the A2C can greatly improve the performance of the model in the long-tail space (\eg the average gain of +2.73\% or +3.05\%), which proves that poor long-tail performance can be solved by aligning the distribution of source and target domains. (2) Compared with the model w/o $L_{DC}^c$ (\eg BM+$L_{DA}$ or ESAM w/o $L_{DC}^c$), the performance of the model w/ $L_{DC}^c$ (\eg ESAM w/o $L_{DC}^p$ or ESAM) is particularly prominent in the hot space (\eg the average gain of +0.6\% or +0.7\%) by adopting $L_{DC}^c$ to optimize the source spatial structure. Besides, $L_{DC}^c$ makes the target spatial structure aligned with better intra-class compactness and inter-class separability, thus improving the lone-tail performance (\eg the average gain of +0.5\% or +1.8\%). (3) The combination of these three constraints (\ie ESAM) yields the best performance, which indicates the necessity of the three regularization terms we design. (4) The ESAM performs better than UIM, which illustrates that ESAM can suppress the "Matthew Effect" to improve long-tail performance.
\subsection{Feature Distribution Visualization}
To illustrate the existence of domain shift and the validity of the proposed ESAM, we randomly select 2,000 items and visualize the features output from $f_d$ using t-SNE \cite{donahue2014decaf}. As shown in Fig.~\ref{tsne}, we can make intuitive observations.  (1) As shown in Fig.~\ref{2D1}, there is a domain gap between distributions of source and target features. This gap makes the model that fits source items cannot be applied in the entire space where has a large number of non-displayed items. (2) We find that the integration of $L_{DA}$ can significantly reduce the disparity between distributions (Fig.~\ref{2D2}), which proves that the correlation between item high-level attributes can reflect the distribution of domain well.  (3) The feature space will have better manifold structure by adopting discriminative clustering. $L_{DC}^c$ enforces the shared feature space with better intra-class compactness and inter-class separability (Fig.~\ref{2D3}). Also $L_{DC}^p$ encourages further more discriminative alignment. Compared to Fig.~\ref{2D3}, $L_{DC}^p$ expands the abscissa to [-60,80] and the ordinate to [-80,80] (Fig.~\ref{2D4}). In a word, ESAM extracts domain-invariant and discriminative features by discriminative domain adaptation, which makes the model robust to long-tail items and can retrieve more personalized and diverse items to users.
\subsection{Score Distribution}
To further illustrate that ESAM can effectively optimize long-tail performance. We randomly select 500 combinations $(\bm{v}_q,\bm{v}_d)$ from source and target domains respectively, to visualize the score distribution. As shown in Fig.~\ref{scoref}, the abscissa is the score which can be calculated as Eq. \ref{consine} and the ordinate is the number of combinations whose scores fall within the interval. As can be seen in Fig.~\ref{s2D1}, the model tends to give extremely low scores for non-displayed items, making these items difficult to be retrieved, while $L_{DA}$ encourages the source and target domains to have the same score distribution, which increases the exposure probability of long-tail items. However, the aligned score distribution is too concentrated (Fig.~\ref{s2D2}), which may result in retrieving irrelevant items. To solve this problem, the proposed discriminative clustering encourages the score distribution to be aligned with better discriminability (Fig.~\ref{s2D3} and Fig.~\ref{s2D4}). As shown in Fig.~\ref{s2D3}, there is an abnormal spike (around score = 0.15) of the target plot, we think this is due to the negative transfer (the positive samples are aligned to negative samples) caused by ignoring the target label during alignment.  As expected, self-training with target reliable pseudo-labels can solve this problem to a certain extent.
\begin{figure}[ht]
    \centering
  \begin{subfigure}[b]{0.235\textwidth}
    \includegraphics[width=\textwidth]{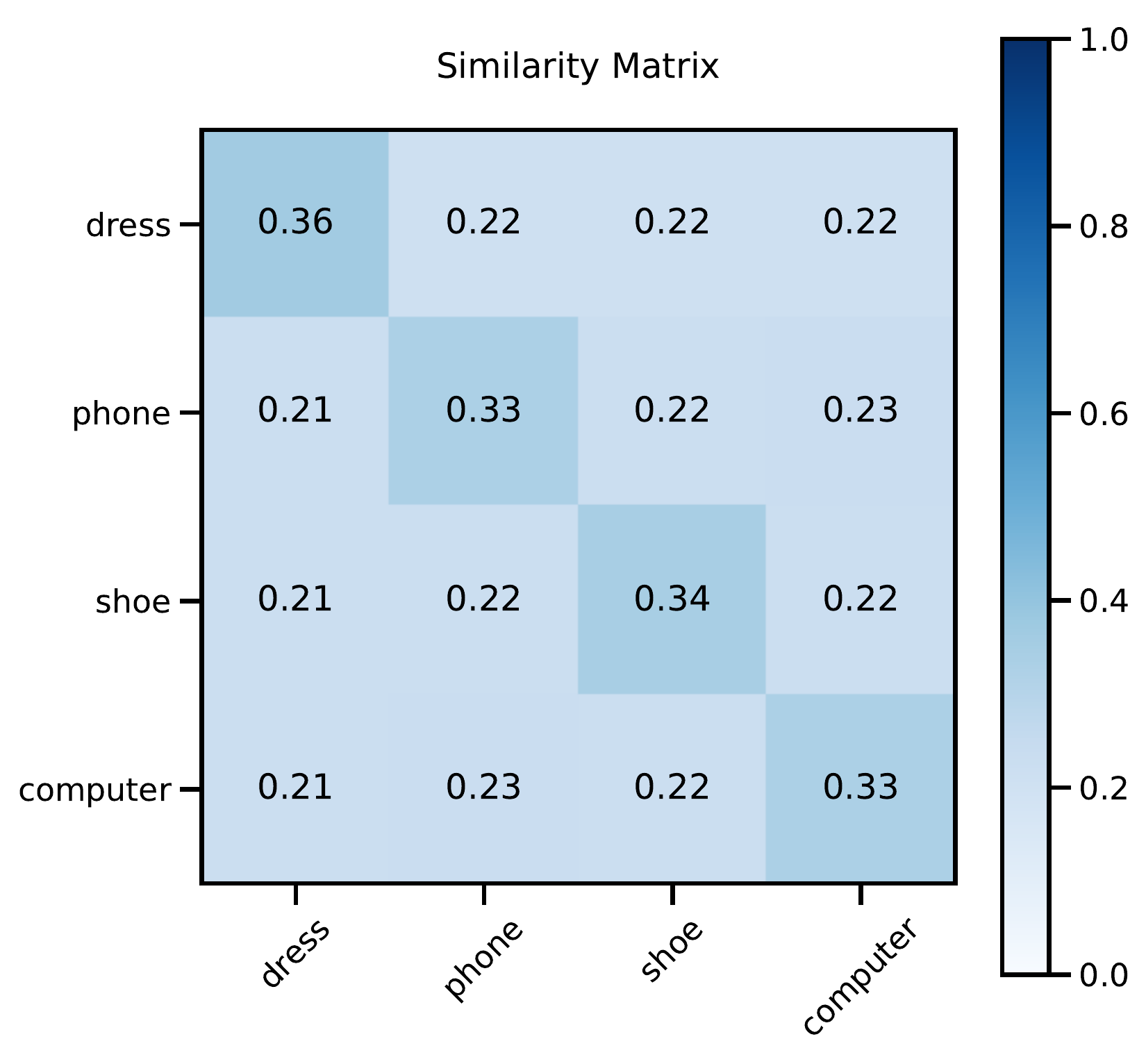}
    \caption{BaseModel}
    \label{si2D1}
  \end{subfigure}
   \begin{subfigure}[b]{0.235\textwidth}
    \includegraphics[width=\textwidth]{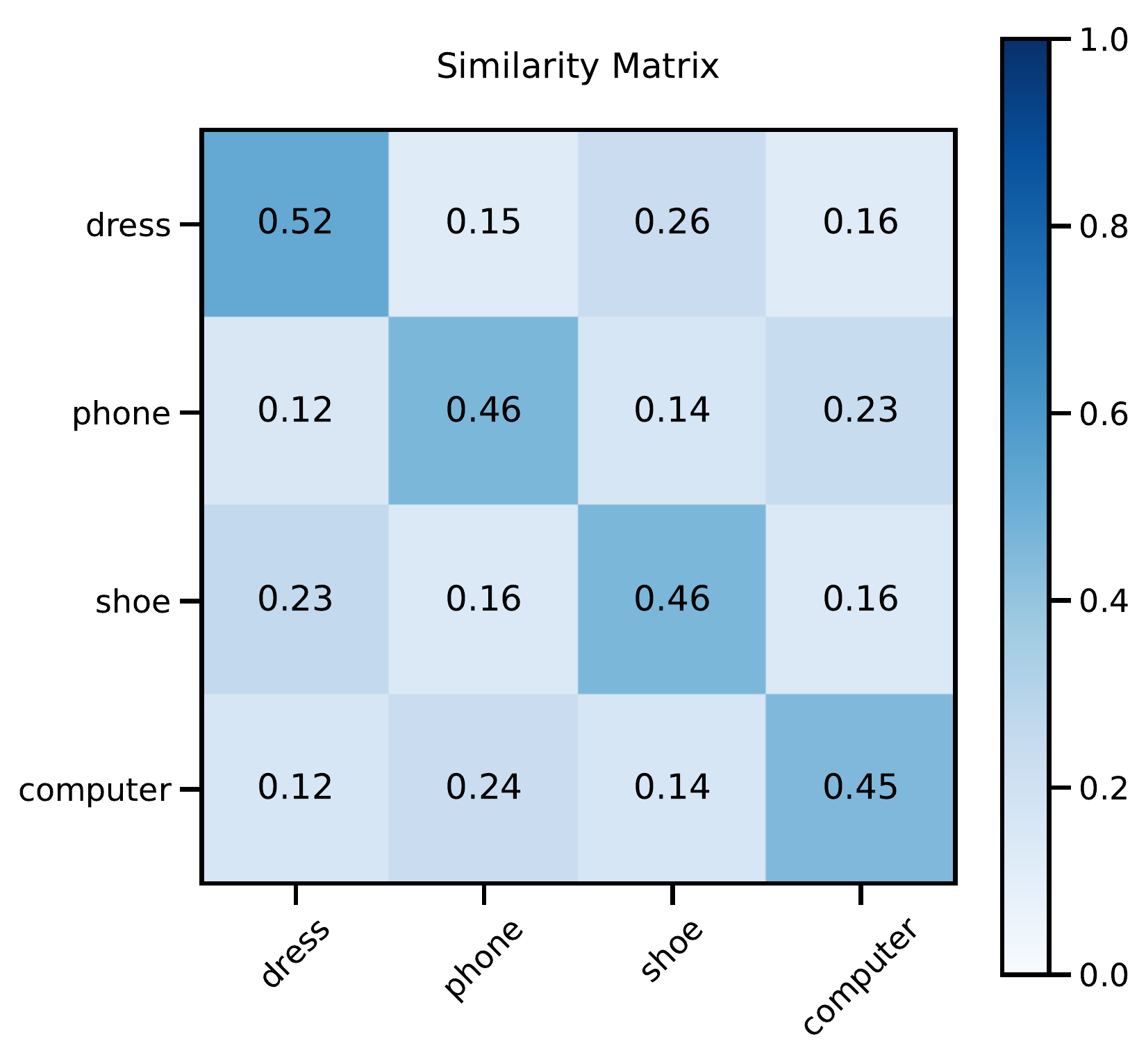}
    \caption{ESAM}
    \label{si2D2}
  \end{subfigure}
\caption{Similarity matrix on Industrial dataset.}
\label{simim}
\end{figure}

\subsection{Similarity Matrix}
To prove that ESAM can learn better item feature representations. We randomly selected $1,000$ items from each category (dress, phone, shoe, and computer) in the entire item space of Industrial dataset. We build two groups of items by equally spliting the item samples (\ie $500$) for each category. For each group, we calculate its category center as the average of the item feature representations. Then, we can obtain the similarity matrix through the cosine similarity for each center pair. As shown in Fig.~\ref{simim}, compared with BaseModel, ESAM has greater intra-class similarity and smaller inter-class similarity, which suggests that ESAM has a more discriminative neighborhood relationship. Besides, the similarity matrix extracted by ESAM can better reflect the real situation, \ie the similarity between dress and shoe is higher than that of dress and phone, which cannot be displayed in the BaseModel.

 \eat{
 divide every 500 items into one group to calculate its category center $
\bm{\alpha}_k^m=\frac{1}{n}\sum_{j=1}^{n}\bm{v}_{d_j^k}$. $k$ represents the category and $m \in \{0,1\}$ represents the m-th center of category $k$.  We can obtain the similarity matrix through the similarity between the category centers. The similarity can be calculated as $simi(k_1.k_2)=\frac{\bm{\alpha}_{k_1}^0\bm{\alpha}_{k_2}^{1\top}}{\|\bm{\alpha}_{k_2}^1\|\cdot\|\bm{\alpha}_{k_1}^0\|}$.
 }

\subsection{Parameter Sensitivity}
\begin{figure}[ht]
    \centering
  \begin{subfigure}[b]{0.46\textwidth}
    \includegraphics[width=\textwidth]{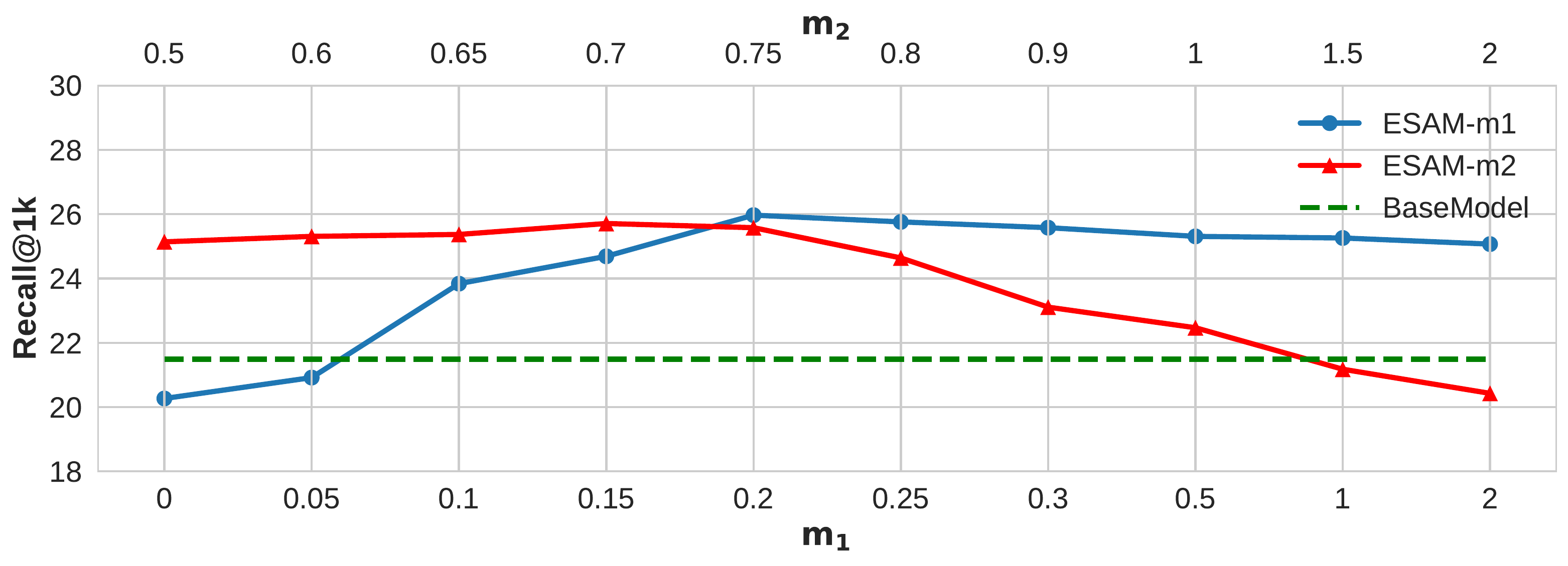}
    \caption{$m_1$ and $m_2$ sensitivity.}
    \label{pa2D1}
  \end{subfigure}
   \begin{subfigure}[b]{0.46\textwidth}
    \includegraphics[width=\textwidth]{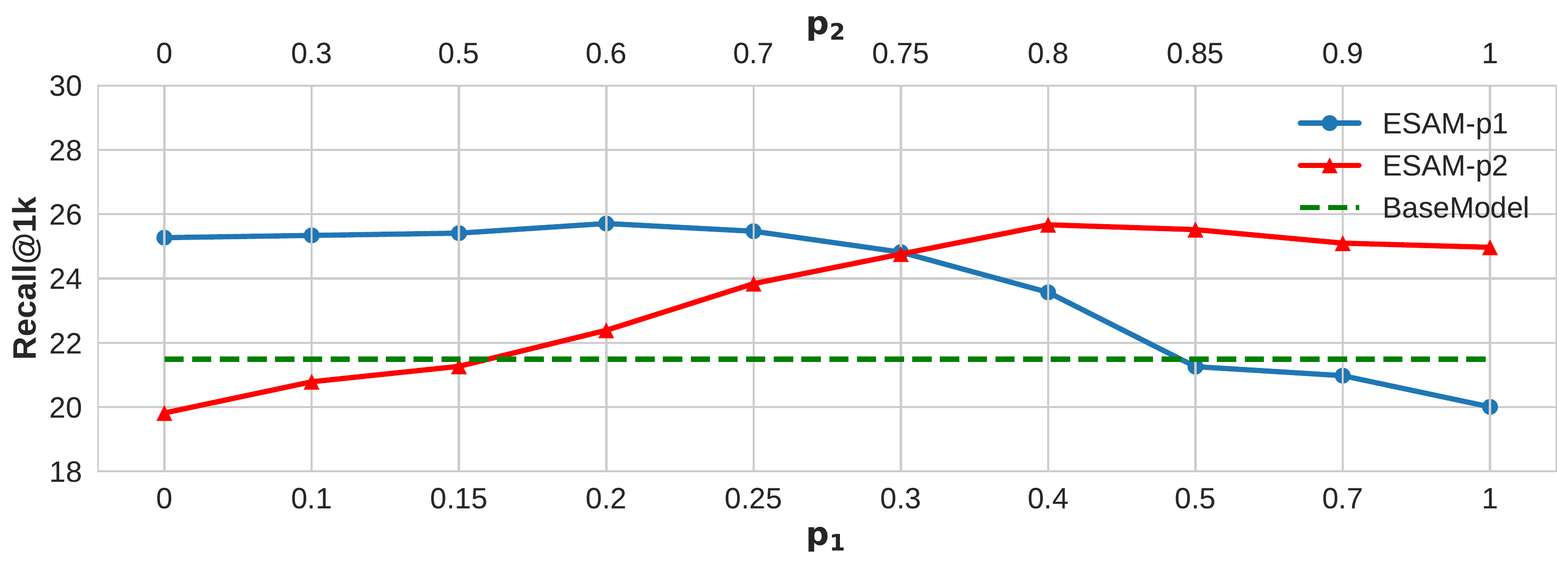}
    \caption{$p_1$ and $p_2$ sensitivity.}
    \label{pa2D2}
  \end{subfigure}
\caption{Parameter sensitivity of $L_{DC}^c$, and confidence thresholds of $L_{DC}^p$ on Industrial dataset. }
\label{para}
\end{figure}

To study the effects of distance constraints $m1$ and $m2$ of $L_{DC}^c$, we vary their value as $m_1 \in \{0,0.05,0.1,0.15,0.2,0.25,0.3,0.5,1,2\}$ and $m_2 \in \{0.5,0.6,0.65,0.7,0.75,0.8,0.9,1,1.5,2\}$. Fig.~\ref{pa2D1} shows bell-shaped curves, which indicates that an appropriately intra-class and inter-class distance constraints can effectively optimize the neighborhood relationship to improve retrieval performance. For the confidence thresholds $p_1$ and $p_2$ of $L_{DC}^p$,  as shown in Fig.~\ref{pa2D2}, we find that assigning pseudo-labels to samples with low confidence ($p_1$ is too large or $p_2$ is too small) for self-training would lead to collapse of the model due to a large number of labels are false ones. While increasing the confidence threshold can improve the performance of the model, which indicates that entropy regularization with a proper constraint is effective to select reliable target samples for better domain adaptation and final performance.

\subsection{Online Experiments}
We deploy the proposed method to Taobao search engine for personalised searching and conduct the online A/B testing experiment.  To make a fair comparison, except for whether the model is integrated with ESAM, other variables are consistent, including user pool,  item pool, feature composition, etc.  Compared with BaseModel, ESAM achieves $0.85\%$ CTR, $0.63\%$ collection rate, $0.24\%$ CVR and $0.4\%$ GMV gain in seven days. The "collection" represents the behavior of adding the item to the cart. Note that, the performance improvement is significant enough for our business due to the search engine in Taobao serves billions of users and generates more than ten million GMV per day. We attribute this to the improved search capability for long-tail items by introducing an unlabeled non-displayed input stream to reduce the distribution discrepancy between displayed and non-displayed items.  In short, the online experiments validate that ESAM can significantly advance the whole search engine. 
\section{Related Work}\label{Related}
\subsection{Neural Network-Based Ranking Model}
Recently, many works have been done to use deep neural networks for many ranking based applications. A significant development in the ranking model is deep learning to rank (LTR)~\cite{cheng2016wide,li2019multi,zhu2019improving,huang2013learning,burges2005learning,severyn2015learning}. To tackle this problem, some methods utilize DA techniques, such as maximum mean discrepancy (MMD) \cite{tran2019domain}, adversarial training \cite{krishnan2018adversarial},  to alleviate inconsistent distributions in the source and target domains. Also, some methods \cite{yuan2019improving,bonner2018causal} introduce an unbiased dataset obtained by an unbiased system (\ie randomly selecting items from the entire item pool to a query) to train an unbiased model. Besides, some methods introduce auxiliary information \cite{yuan2019darec,yuan2019improving} or auxiliary domains \cite{gao2019cross,man2017cross} to obtain more long-tail information. Unlike previous approaches, ESAM joints domain adaptation and  non-displayed items to improve long-tail performance without any auxiliary information and auxiliary domains. Moreover, we design a novel DA technique, named attribute correlation alignment, which regards the correlation between item high-level attributes as knowledge to transfer.

\subsection{Discriminative Domain Adaption}
Domain adaptation, which transfers knowledge from a large number of labeled source samples to target samples with missing or limited labels for target performance improvement, has been widely studied recently. These DA methods learn the domain-invariant feature by embedding the adaptation layer for moment matching, such as maximum mean discrepancy (MMD) \cite{chendeep}, correlation alignment (CORAL) \cite{sun2016deep, chen2019selective} and center moment discrepancy (CMD) \cite{zellinger2017central}, or integrating a domain discriminator for adversarial training, such as domain adversarial neural network (DANN) \cite{ganin2016domain} and adversarial discriminative domain adaptation (ADDA) \cite{tzeng2017adversarial}. Some previous works pay efforts to learn more discriminative features for performance improvement, \eg the contrastive loss \cite{hadsell2006dimensionality} and the center loss \cite{liu2016large}. These methods have been adopted in many applications, such as face recognition and person re-identification, etc.  Inspired by these methods, we propose to perform domain adaptation to lean item features with better discriminative power in entire space.

\section{Conclusion}\label{conclusion}
In this paper, we propose ESAM to improve long-tail performance with discriminative domain adaptation by introducing non-displayed items. To the best of our knowledge, this is the first work to adopt domain adaptation with non-diplayed items for ranking model. It is worth mentioning that ESAM is a general framework, which can be easily integrated into many existing ranking models. The offline experiments on two public datasets and a Taobao industrial dataset prove that ESAM can be integrated into the existing SOTA baselines to improve retrieval performance, especially in the long-tail space. Online experiments further demonstrate the superiority of ESAM at Taobao search engine. Furthermore, we also verify the necessity of each constraint by ablation studies. 
\begin{acks}
    This work was supported by Alibaba Group through Alibaba Innovative Research Program and National Natural Science Foundation of China (No.~61872278). Chenliang Li is the corresponding author.
\end{acks}

\bibliographystyle{ACM-Reference-Format}
\bibliography{sample-base}

\end{document}